\documentclass[prd,superscriptaddress,floatfix,nofootinbib,twocolumn,aps]{revtex4-1}
\usepackage{graphicx,amssymb,amsmath,bm,dcolumn,xcolor,mathtools}
\usepackage[caption=false]{subfig}
\usepackage{threeparttable}
\usepackage{comment}
\usepackage{xfrac}
\usepackage{appendix}
\usepackage{enumitem}
\usepackage[normalem]{ulem}
\usepackage{soul}
\usepackage{multirow}
\begin{document}

\title{Searching for Scalar Ultralight Dark Matter with Optical Fibers}

\author{J. Manley}
\affiliation{Department of Electrical and Computer Engineering, University of Delaware, Newark, DE 19716, USA}

\author{R. Stump}
\affiliation{Department of Electrical and Computer Engineering, University of Delaware, Newark, DE 19716, USA}

\author{R. Petery}
\affiliation{Department of Electrical and Computer Engineering, University of Delaware, Newark, DE 19716, USA}

\author{S. Singh}
\affiliation{Department of Electrical and Computer Engineering, University of Delaware, Newark, DE 19716, USA}

\date{\today}

\begin{abstract}
We consider optical fibers as detectors for scalar ultralight dark matter (UDM) and propose using a fiber-based interferometer to search for scalar UDM with particle mass in the range $10^{-17} - 10^{-13}$ eV/$c^2$ $\left(10^{-3}- 10 \text{ Hz}\right)$. Composed of a solid core and a hollow core fiber, the proposed detector would be sensitive to relative oscillations in the fibers' refractive indices due to scalar UDM-induced modulations in the fine-structure constant $\alpha$. We predict that, implementing detector arrays or cryogenic cooling, the proposed optical fiber-based scalar UDM search has the potential to reach new regions of the parameter space. Such a search would be particularly well-suited to probe for a Solar halo of dark matter with a sensitivity exceeding that of previous DM searches over the particle mass range $7\times 10^{-17} - 2\times 10^{-14}$ eV/$c^2$. 
\end{abstract}

\maketitle

\section{Introduction}
Dark matter (DM) is a substance of unknown composition that accounts for 85\% of the matter in the Universe. It is the dominant component of matter in galaxies~\cite{particle2020review}, and its existence is inferred through its gravitational effects on normal matter. The lack of non-gravitational observations permits a wide variety of viable DM candidates, along with interactions with Standard Model particles and fields~\cite{battaglieri2017us, antypas2022new}.

\begin{figure}[b]
	\centering
	\includegraphics[width=.98 \columnwidth,clip, trim= 0.1in 1.5in 2in 1.3in]{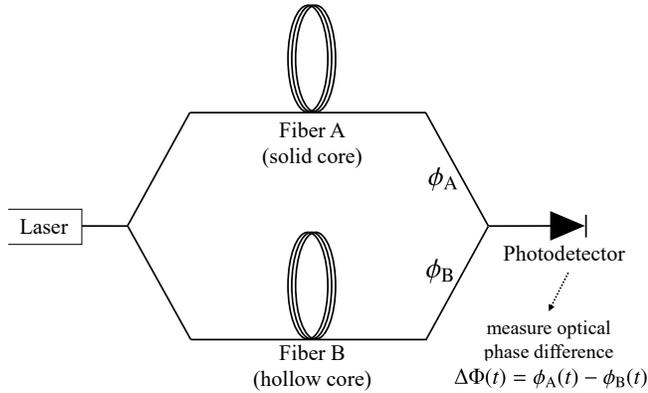}
	\caption{\textbf{Conceptual model for an optical fiber-based detector:} Relative oscillations in the refractive indices of two fibers can be detected using an optical-path-length-balanced Mach-Zehnder interferometer. Scalar UDM affects the refractive index of the solid core fiber, shifting the optical phase and resulting in an oscillation in output optical power at the UDM frequency. While the noise analysis in this paper assumes a more complex setup (see Fig. \ref{fig:schematic}), which leads to common-mode rejection of various technical noise, the signal can be well-modeled by the setup above.}
	\label{fig:balancedSchematic}
\end{figure}

We consider a scalar ultralight dark matter (UDM) scenario, where dark matter is entirely composed of scalar particles with mass $m_\text{DM}\lesssim 10$ eV/$c^2$ that would act as a coherently oscillating scalar field around earth due to its large number density. Through non-gravitational interactions with normal matter, scalar UDM induces an effective oscillation in fundamental constants~\cite{damour1994string}. Specifically, we consider oscillation of the fine-structure constant $\alpha$, an effect that can be searched for with a wide variety of detector types~\cite{antypas2022new} including atomic experiments~\cite{van2015search,hees2016searching,beloy2021frequency,filzinger2023improved,sherrill2023analysis,campbell2021searching}, optical cavities \cite{geraci2019searching,kennedy2020precision,savalle2021searching}, mechanical systems~\cite{arvanitaki2015searching,manley2020searching}, and gravitational wave detectors~\cite{vermeulen2021direct,aiello2021constraints,branca2017search}. 

Here we propose a detector that would use optical fibers to search for scalar UDM-induced oscillations in $\alpha$, which would produce a measurable oscillation of optical refractive indices. Such a detector would probe for UDM by comparing the refractive indices of two different types of fibers, solid core and hollow core, which would be differentially affected by oscillations in $\alpha$. %Our work shows that optical fibers can be repurposed as laboratory-scale dark matter detectors, potentially achieving sensitivities to scalar UDM that exceed the current constraints at sub-Hz frequencies.

Due to ease of manufacturing coupled with extremely low optical loss, optical fibers are low cost and large channel capacity information transmission devices. These properties make them ubiquitous in long distance classical and quantum communication networks. Technical advances in ultra low loss fibers are being accelerated due demands in disparate fields such as high-volume online data transmission or quantum cryptography. In a parallel development, photonic crystal fibers offer a novel route to efficiently transmit high power without losses associated with material non-linearity \cite{russell2006photonic}. When combined with advances in cryogenics fueled by quantum computing, existing and near-term fiber technology offers a promising table-top route to search for dark matter, achieving sensitivities to scalar UDM that exceed the current constraints over a wide range of sub-Hz frequencies.

This paper is organized as follows: In Section \ref{sec:Theory}, we provide a brief background of scalar UDM and the signal it produces. In Section \ref{sec:fiberDetector}, we introduce the concept of an optical fiber-based UDM detector. In Section \ref{sec:noiseSources} we present a model for the noise sources that would limit the detector's sensitivity and in Section \ref{sec:dmin} we use this noise model to calculate the minimum detectable coupling strength, which is evaluated considering both Galactic halo and Solar halo UDM scenarios. Additional details can be found in the Appendices, including derivations of expressions in the main text and extended discussion of the noise models used to characterize the prospective UDM detector.

%-------------------------------------------------------------
%-------------------------------------------------------------
\section{Scalar UDM} \label{sec:Theory}

%-------------------------------------------------------------
%-------------------------------------------------------------
For the range of particle masses considered in this work, UDM can be considered a coherent and approximately spatially uniform %\footnote{For particle masses below $10^{-13}$ eV/$c^2$, the de Broglie wavelength exceeds $10^7$ km.} 
background field on Earth. This field is nearly monochromatic, oscillating at the Compton frequency $f_\text{DM}=m_\text{DM}c^2/h$. Through Doppler broadening, the velocity dispersion of DM gives the field a finite coherence time $\tau_\text{DM}$. Over timescales less than $\tau_\text{DM}$ the field can be expressed as a sinusoid,
\begin{equation} \label{scalarUDMField}
\varphi(t) \approx \varphi_0\cos \left(2\pi f_\text{DM} t + \theta_\text{DM}\right),
\end{equation} 
whose amplitude $\varphi_0$ is a stochastic quantity described by a Rayleigh distribution~\cite{centers2021stochastic}. The field amplitude's root-mean-square value $\varphi_\text{rms} =\sqrt{\frac{2 G \rho_\text{DM}}{\pi c^2 {f_\text{DM}}^2}}$ is determined by the local dark matter density $\rho_\text{DM}$. The values of $\tau_\text{DM}$ and $\rho_\text{DM}$ depend on the DM halo model being considered. For example, in this work we consider two halo models, (1) a standard, smooth Galactic halo where $\rho_{\text{DM}}\approx 0.4$ GeV/cm\textsuperscript{3}~\cite{particle2020review} and $\tau_\text{DM}\approx 10^6 {\omega_\text{DM}}^{-1}$~\cite{derevianko2018detecting}, as well as (2) a local UDM halo centered on the Sun, which would have a greater particle density and coherence time (see Section \ref{sec:dmin} for more details). The unknown phase $\theta_\text{DM}$ has a flat distribution from $0$ to $2\pi$~\cite{centers2021stochastic}. While these temporal and spatial field properties are general to multiple types of UDM, the experimental signature of scalar UDM depends on its specific interactions with Standard Model particles. 

We assume scalar UDM couples linearly to the electromagnetic field tensor $F^{\mu\nu}$. Such an interaction is described by the Lagrangian density term~\cite{damour2010equivalence}
\begin{equation}\label{scalarInteractionLagrangian}
\mathcal{L}_\text{int} =  d_e \varphi(t) \frac{e^2 c}{16\pi \hbar \alpha_0}F_{\mu\nu}F^{\mu\nu}
\end{equation}
where $e$ is the elementary electric charge, $c$ is the speed of light, $\hbar$ is the reduced Planck constant, and $\alpha_0$ is the fine-structure constant (in the absence of scalar UDM). The strength of the interaction is parameterized with the dimensionless coupling strength $d_e$. 

Generally, interactions between scalar UDM fields and normal matter can be modeled as variations of fundamental constants~\cite{damour1994string}, such as the electron mass $m_e$ or fine-structure constant $\alpha$, where the value of a fundamental constant at any point in space or time depends on the local value of the scalar UDM field. In the case of Eq. \ref{scalarInteractionLagrangian}, scalar UDM causes fractional fluctuations in the fine-structure constant given by
\begin{equation}  \label{alphaModulation}
\frac{\delta \alpha}{\alpha_0}(t)=d_{e} \varphi(t).
\end{equation}
Oscillations in $\alpha$ produce measurable effects such as modulation of atomic energy levels~\cite{arvanitaki2015searching}, material refractive indices~\cite{braxmaier2001proposed}, and mechanical strains~\cite{stadnik2015searching,arvanitaki2016sound}.

Scalar UDM-induced oscillations in the fine-structure constant lead to a material-dependent oscillation of optical refractive indices, and the proposed fiber-based detector would be sensitive to scalar UDM primarily through its modulation of an optical fiber's refractive index. The refractive index of an optical material generally depends on the fine-structure constant~\cite{grote2019novel,braxmaier2001proposed}. For small fluctuations in $\alpha$, the resulting fractional fluctuation in refractive index is
\begin{equation} \label{indexModulation}
    \frac{\delta n}{n_0} = \epsilon_{n\alpha}\frac{\delta \alpha }{\alpha_0},
\end{equation}
where $\epsilon_{n\alpha}\equiv \frac{\alpha_0}{n_0}\frac{\partial n}{\partial \alpha}\big|_{\alpha=\alpha_0}$. While the coefficient $\epsilon_{n\alpha}$ may not be directly measurable, it can be related to a material's optical dispersion as~\cite{grote2019novel,braxmaier2001proposed}
\begin{equation} 
\epsilon_{n\alpha}=-2 \epsilon_{n\omega_\text{L}},
\label{eq:epsilon_nalpha}
\end{equation}
where $\epsilon_{n\omega_\text{L}}\equiv \frac{\omega_\text{L}}{n_0}\frac{\partial n}{\partial \omega_\text{L}}\big|_{\omega_\text{L}=\omega_{0,\text{L}}}$, $\omega_\text{L}$ is the optical angular frequency, and $\omega_{0,\text{L}}$ is the angular frequency of the laser. Details of the model used to derive Eq. \ref{eq:epsilon_nalpha} are given in Appendix \ref{app:DMindex}.

\section{Optical fibers as DM detectors} \label{sec:fiberDetector}
Here we propose to search for scalar UDM by comparing the refractive indices of two different types of fibers, solid core and hollow core fibers, which are differentially affected by oscillations in the fine-structure constant. In solid core fibers, the optical mode is contained within silica ($n_\text{A} \approx n_\text{silica}\approx 1.5$), for which $\epsilon_{n\omega_\text{L},\text{silica}}= 0.013$ at an optical wavelength of $1550$ nm (see Appendix \ref{app:DMindex}). In hollow core fibers, the optical mode is mostly contained within air (or vacuum) ($n_\text{B} \approx n_\text{air}\approx1$~\cite{shi2021thinly}), for which $\epsilon_{n\omega_\text{L},\text{air}}\ll \epsilon_{n\omega_\text{L},\text{silica}}$~\cite{bradley2019antiresonant,jasion2021recent}. 

A simple model for a fiber-based scalar UDM detector is an optical-path-length-balanced Mach-Zehnder interferometer, as depicted in Fig. \ref{fig:balancedSchematic}, where a solid core fiber and a hollow core fiber each constitute an interferometer arm. By balancing the optical path lengths of each fiber, some common mode noise sources can be suppressed, and oscillations in the relative optical lengths of the fibers can be measured interferometrically using a laser with central wavelength $\lambda_{0,\text{L}}=1550$ nm and average power $\sim$1 mW. Small fluctuations in $\alpha$ due to scalar UDM result in a phase difference signal with magnitude
\begin{equation}\label{dmPhaseSignal}
\Delta\Phi^\text{DM}=  \omega_{0,\text{L}}\tau_0  \epsilon_{n\omega_\text{L},\text{silica}}\frac{\delta \alpha}{\alpha_0},
\end{equation}
where $\tau_0=n_{0}L_{0}/c$ is the time delay. A detailed derivation of the output phase difference, including the UDM-induced phase difference given in Eq. \ref{dmPhaseSignal}, is given in Appendix \ref{phaseDifferenceAppendix}. 

Appendix \ref{phaseDifferenceAppendix} also introduces a modified version of the interferometric setup (see Fig. \ref{fig:balancedSchematic}), which enables the cancellation of several phase noise sources.
As detailed derivations in Appendix \ref{phaseDifferenceAppendix} and \ref{app:laserNoise} show, the detector geometry displayed in Fig. \ref{fig:schematic} has a sensitivity that is consistent with a balanced interferometer (represented in Fig. \ref{fig:balancedSchematic}), while being immune to laser frequency noise. We note that the quantitative results presented below in Figs. \ref{fig:phaseNoisePlot} and \ref{fig:dminPlots} are for the setup shown in Fig. \ref{fig:schematic}.

%While the detector's ability to detect scalar UDM relies on a comparison of optical refractive indices, it is worth mentioning that 
Scalar UDM also produces a strain signal~\cite{stadnik2015searching,arvanitaki2016sound} that affects the experiment via a coherent oscillation in the lengths of the fibers and the laser cavity. Because both fibers are primarily composed of the same materials, the strain is common to both interferometer arms and does not produce a measurable phase difference (also, the material dependence of the strain would be a higher-order effect~\cite{pavsteka2019material}). However, straining the laser cavity shifts the laser's central wavelength, producing an optical path length imbalance due to differences of optical dispersion between the fibers. Therefore, while the presence of scalar UDM would ultimately be inferred from relative oscillations in the fibers' refractive indices, it is worth noting that this signal includes contributions from both direct modulation of the solid core fiber's refractive index as well as the strain in the laser cavity (via optical dispersion in the fibers). Incidentally, for the case of scalar UDM coupling to the electron mass, this effect cancels the signal produced through direct refractive index modulation of the fibers (see Appendix \ref{app:DMindex}). For this reason, the detector is only sensitive to oscillations in the fine-structure constant. We note that a different combination of fiber-based interferometry and laser stabilization might enable measurement of a differential length change effect, or access to the electron mass coupling to UDM. More importantly, our analysis highlights the importance of including the effect of the omnipresent UDM signal on the experimental apparatus beyond the interferometer.

\begin{figure}[ht]
	\centering
	\includegraphics[width=0.98 \columnwidth]{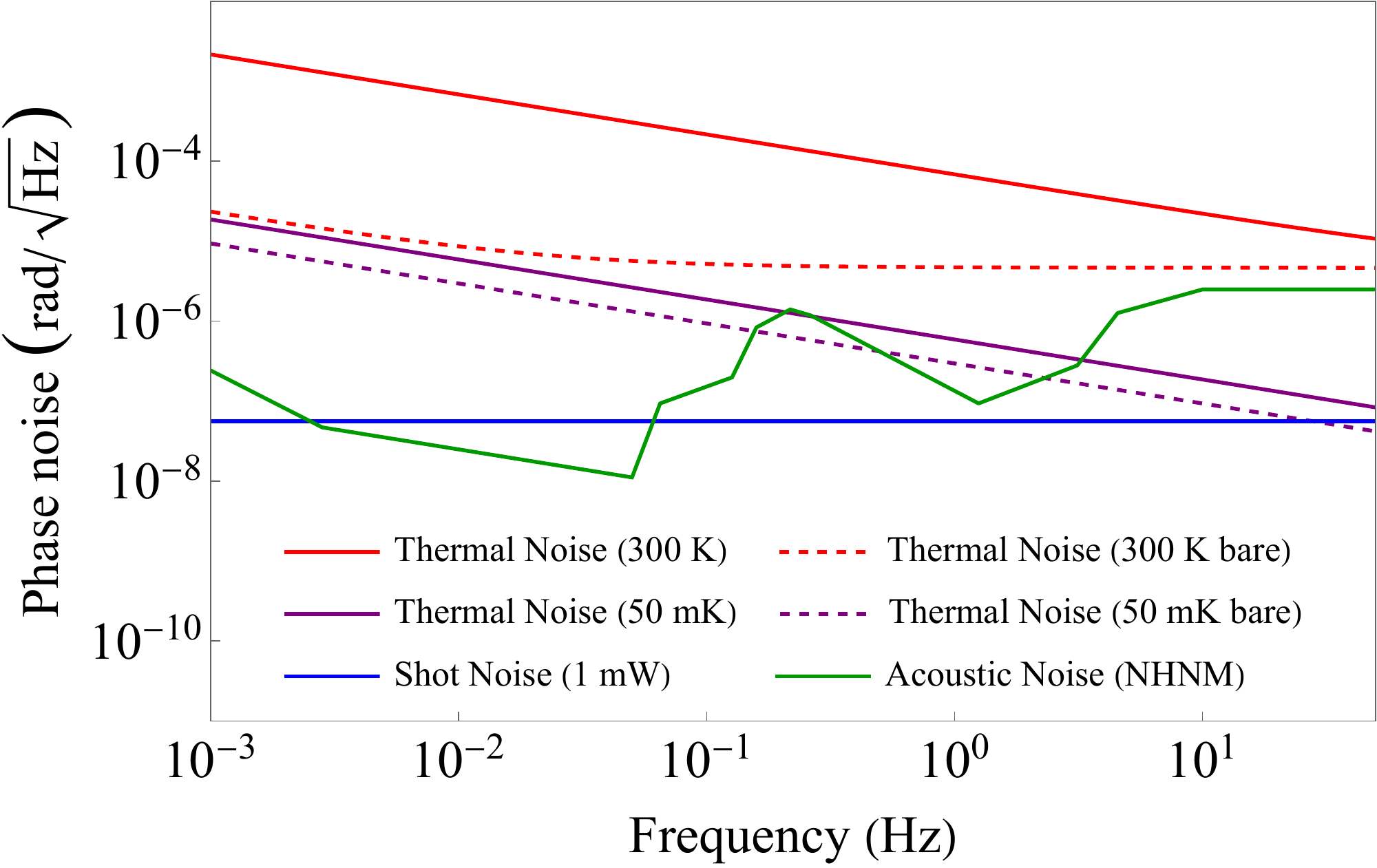}
	\caption{\textbf{Estimated phase noise for the fiber-based detector:} Contributions are included from each of the noise sources detailed in Section \ref{sec:noiseSources}. Several thermal noise curves are included, corresponding to fibers with (solid lines) and without (dashed lines) polymer coatings at temperatures of $300$ K and $50$ mK.}
	\label{fig:phaseNoisePlot}
\end{figure}

%-------------------------------------------------------------
%-------------------------------------------------------------
\section{Noise}  \label{sec:noiseSources}
%-------------------------------------------------------------
%-------------------------------------------------------------
Various noise sources have the potential to overshadow the DM signal $\Delta\Phi^\text{DM}$. The dominant noise sources to be considered when designing a fiber-based UDM detector are thermal noise and acoustic noise in the fibers themselves, noise from the laser, and photon shot noise.  The effect of each noise source is quantified by a one-sided phase noise power spectral density (PSD) $S_{\Delta \Phi}(f)$, and plotted in Fig. \ref{fig:phaseNoisePlot}.

Thermomechanical noise in the fibers will likely set the ultimate limit to the detector's sensitivity. Fiber ``thermomechanical noise" $S_{\delta\phi}^\text{TM}$ refers to the optical phase fluctuations $\delta \phi$ that are induced by spontaneous thermal fluctuations of a fiber's length due to internal friction ~\cite{duan2012general}. Fluctuation-dissipation theorem provides an accurate model for thermomechanical noise in fibers, resulting in a phase noise PSD given by~\cite{duan2010intrinsic} 
\begin{equation} \label{thermomechanicalNoise}
S_{\delta\phi,i}^\text{TM}(f) = \left(\frac{2\pi n_{0,i}}{{\lambda_{0,\text{L}}}}\right)^2 \frac{2k_\text{B} T L_{0,i} \xi_i}{3\pi E_i A_i}\frac{1}{f},
\end{equation}
where  $T$ is the temperature, $L_{0,i}$ is the physical length, $\xi_i$ is the mechanical loss tangent, $E_i$ is the Young's modulus, and $A_i$ is the cross-sectional area of Fiber $i$. The resulting differential phase noise in the interferometer is the sum of contributions from both fibers A and B: $S_{\Delta\Phi}^\text{TM} = S_{\delta\phi,\text{A}}^\text{TM}+S_{\delta\phi,\text{B}}^\text{TM}$. Several experiments have directly measured thermomechanical noise in single mode fibers~\cite{bartolo2012thermal,dong2016observation}, and the thermomechanical noise model described by Eq. \ref{thermomechanicalNoise} has been verified down to frequencies as low as 0.05 Hz~\cite{huang2019all}.

As a method for reducing thermomechanical noise we propose reducing the thickness of the fibers' polymer coatings. Mechanical loss in fibers likely comes primarily from the polymer coatings that protect the silica cladding, and there is some evidence suggesting that reducing the thickness of these coatings reduces the thermomechanical noise floor in optical fibers~\cite{dong2016observation}. We include thermal noise estimates in Fig. \ref{fig:phaseNoisePlot} for bare (without coating) fibers as a benchmark approximation for thinly-coated fibers, noting that while fiber coatings can be made quite thin~\cite{shi2021thinly}, bare fibers may be difficult to manufacture and handle due to fragility. Appendix \ref{sectionThermalNoise} has a detailed discussion of mechanical dissipation in optical fibers, including the values of the parameters used in the thermal noise estimates for Fig. \ref{fig:phaseNoisePlot}.

Another way to reduce thermal noise is to implement cryogenic cooling. Figure \ref{fig:phaseNoisePlot} illustrates the impact on thermal noise of lowering the temperature from room temperature ($300$ K) to cryogenic temperature ($50$ mK). It is clear from the figure that the benefit of removing the coatings diminishes at lower temperature, where the intrinsic mechanical loss tangent $\xi$ of fused silica increases beyond its room-temperature value. 

To achieve a thermally-limited detection sensitivity, steps need to be taken to shield the experiment from acoustic noise in the fibers. Mechanical vibrations, which may originate from a variety of sources such as human activity, natural seismic activity, or local weather, often pose a challenge for terrestrial experiments at low frequencies. Vibration-induced strains in optical fibers can be suppressed by using low vibration sensitivity fiber spools~\cite{li2011low} and by co-winding the fibers on the same spool to correlate the acoustic noise between them~\cite{bartolo2012thermal,dong2016observation}. The remaining effects then come from differences in the strain optic effect between the two fibers. To estimate the effects of mechanical vibrations, we use the USGS New High Noise Model (NHNM) for seismic noise~\cite{peterson1993observations}. By co-winding both fibers on a single vibration-insensitive spool, acoustic noise at the level predicted by the NHNM can be reduced to a level comparable to that of thermal noise in this experiment, as can be seen in Fig. \ref{fig:phaseNoisePlot}. 

\begin{figure*}
    \centering
    \subfloat{{\includegraphics[width = 245.0pt]{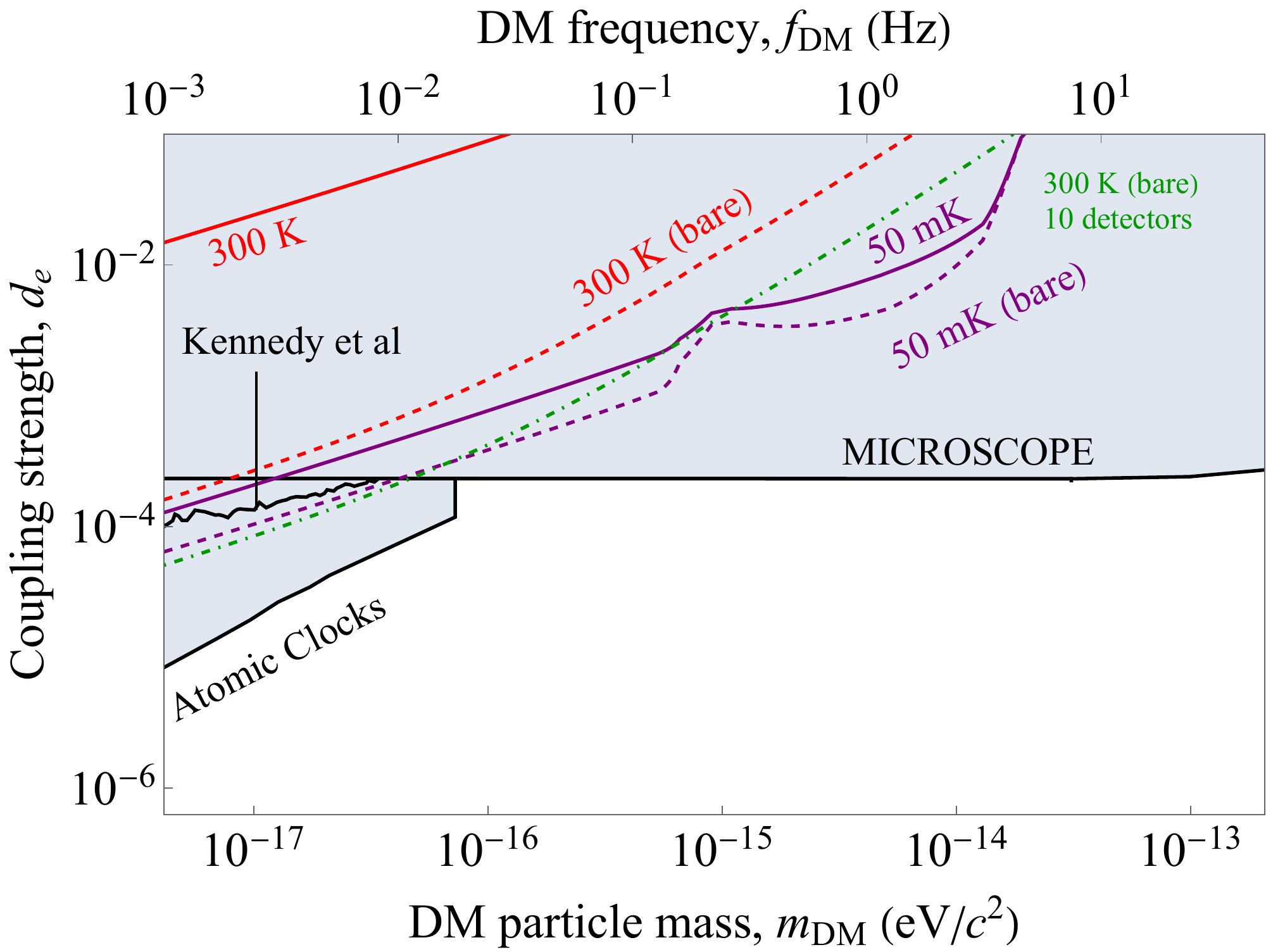}}}
	\qquad
    \subfloat{{\includegraphics[width=245.0pt]{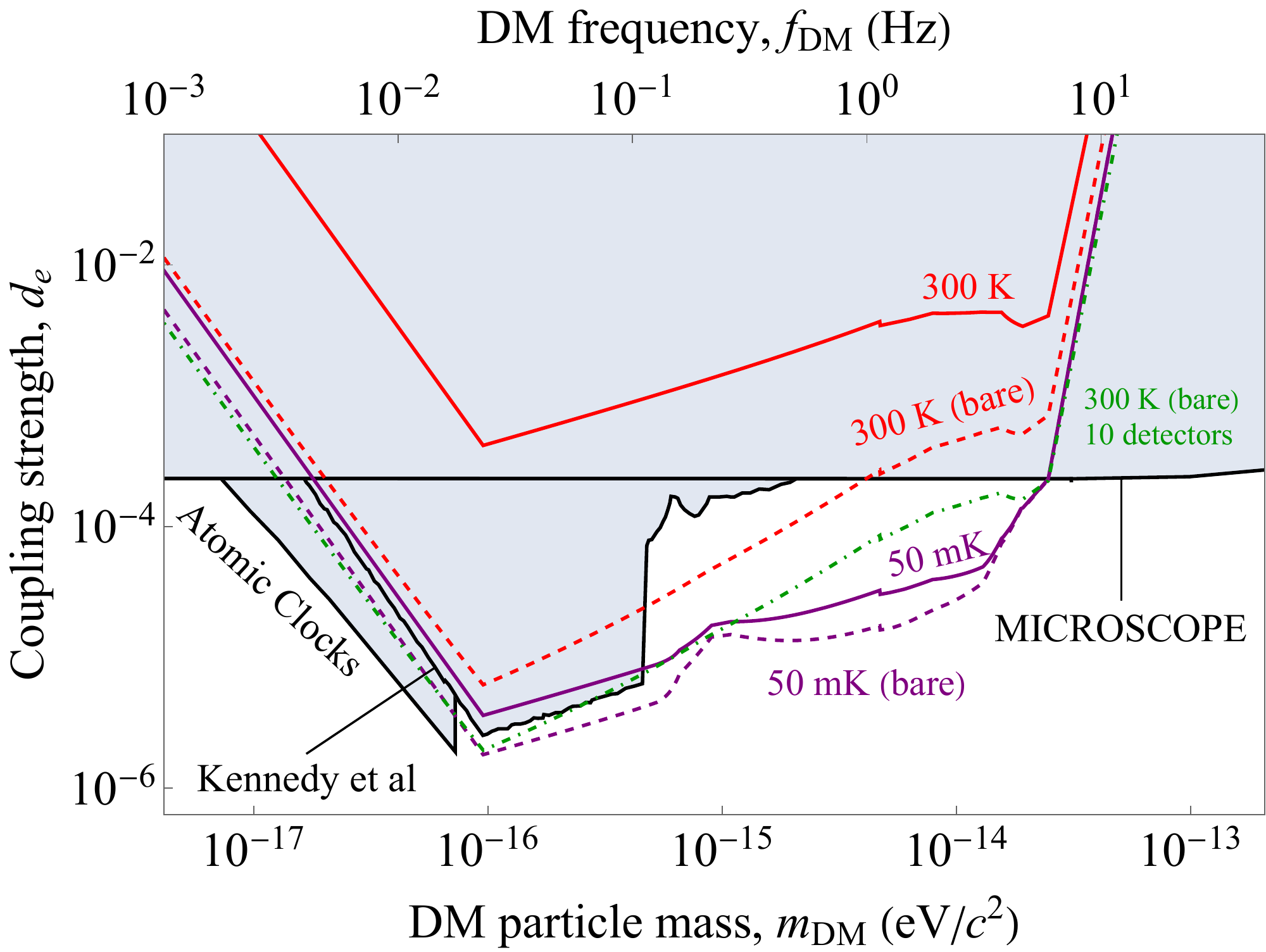}}}
	\caption{\textbf{Projected constraints on $d_e$ from optical fibers, for Galactic (left) and Solar (right) DM halos:} Curves are included for fiber-detectors operating at temperatures of $300$ K and $50$ mK, with (solid) and without (dashed) polymer coatings, for a measurement time $\tau_\text{exp}=1$ year. The fiber lengths are chosen to be $L_{0,\text{A}}=10$ km and $L_{0,\text{B}}\approx 15$ km. The detectors are primarily limited by thermal noise; cryogenic detectors start to become limited by acoustic noise at frequencies above $\sim 10^{-1}$ Hz. Also included are projected constraints from an array of $10$ detectors, using bare fiber at room temperature. The strongest experimental constraints in this mass range come from MICROSCOPE~\cite{berge2018microscope}, Kennedy et al~\cite{kennedy2020precision}, and atomic clock comparisons (here included as a rough, combined constraint from Filzinger et al~\cite{filzinger2023improved} and Sherrill et al~\cite{sherrill2023analysis}).}
	\label{fig:dminPlots}
\end{figure*}

The final noise source included in Fig. \ref{fig:phaseNoisePlot} is photon shot noise, which can be limited to a subdominant level by ensuring sufficient output optical power despite propagation losses in the fibers. Recent advances in hollow core fiber technology have yielded hollow core fibers with a propagation loss of $0.17$ dB/km at $\lambda_{0,\text{L}}=1550$ nm~\cite{jasion20220}, reaching the level of low-loss solid core fibers ($0.18$ dB/km for SMF28~\cite{smf28}). We find that shot noise in this experiment can be limited to below the thermal noise floor at cryogenic temperatures with an input laser power of $P_\text{L}\sim1$ mW.  The laser shot noise can be further reduced by employing quantum optical techniques such as squeezing. 

A more detailed discussion of these noise sources can be found in Appendix \ref{app:noiseSources}, along with discussion of techniques to address frequency and intensity noise in the laser. In addition, Ref. \cite{hilweg2022limits} also has a comprehensive review on the noise performance of optical fibers in interferometric setups for precision measurements.

\section{Minimum detectable coupling strength} \label{sec:dmin}
The smallest signal a detector is capable of measuring can be inferred from the modeled noise floor, and is improved upon by increasing the duration of the measurement $\tau_\text{exp}$. Here, the minimum detectable signal strength $\Delta \Phi_\text{min}^\text{DM}$ is defined as the signal strength needed to achieve a unity signal to noise ratio. For measurement times less than the DM coherence time ($\tau_\text{exp}\lesssim \tau_\text{DM}$), the signal can be considered coherent and stochastic, such that $\Delta \Phi_\text{min}^\text{DM}\propto 3{\tau_\text{exp}}^{-1/2}$~\cite{centers2021stochastic}. For significantly longer measurement times ($\tau_\text{exp}\gg \tau_\text{DM}$),  $\Delta \Phi_\text{min}^\text{DM}\propto \left(\tau_\text{DM}\tau_\text{exp}\right)^{-1/4}$~\cite{budker2014proposal}. Here, we combine both regimes into a single smooth function (see Appendix \ref{app:SignalAnalysis} for more details) to approximate the minimum detectable signal: 
\begin{equation} \label{phiminEquation}
\Delta \Phi_\text{min}^\text{DM}\approx \sqrt{2S_{\Delta \Phi}^\text{noise}} 
\left[3{\tau_\text{exp}}^{-1/2}+\left(\tau_\text{DM}\tau_\text{exp}\right)^{-1/4}\right].
\end{equation}
%\textcolor{blue}{Appendix \ref{app:SignalAnalysis} discusses the viability of Eq. \ref{phiminEquation}, along with other approximations in deriving the minimum detectable signal.}

The utility of a potential UDM experiment is typically evaluated by its projected minimum detectable coupling strength $d_{e,\text{min}}$, a quantity that depends on the astrophysical DM model being considered. The minimum detectable coupling strength $d_{e,\text{min}}$ for the detector can be related to the signal strength via Eq. \ref{dmPhaseSignal} as
\begin{equation} \label{dminEquation}
d_{e,\text{min}} = \frac{\Delta\Phi_\text{min}^\text{DM}}{\omega_{0,\text{L}} \tau_0 \epsilon_{n\omega_\text{L},\text{silica}}\varphi_\text{rms}}.
\end{equation}

The minimum detectable coupling strength $d_{e,\text{min}}$ depends on the local energy density $\rho_\text{DM}$ and coherence time $\tau_\text{DM}$ of the DM; here we consider two different astrophysical DM models for these parameters. The spatial distribution of dark matter across the Milky Way is typically estimated with the Standard Halo Model, which assumes a gravitationally-stable and spherically-symmetric density profile that decreases with distance from the center of the Galaxy $\rho_\text{DM}(r)\propto r^{-2}$ and a Maxwellian velocity distribution~\cite{particle2020review}. 
 
The most commonly used model for UDM in direct detection experiments assumes a smooth Galactic halo, lacking any noticeable substructure, where the DM density and coherence time at Earth take on values determined by the average density and velocity dispersion within the Solar Neighborhood: $\rho_{\text{DM}}\approx 0.4$ GeV/cm\textsuperscript{3}~\cite{particle2020review} and $\tau_\text{DM}\approx 10^6 {\omega_\text{DM}}^{-1}$~\cite{derevianko2018detecting}. The projected constraints from optical fibers, using a Galactic halo model for DM, are plotted in Fig. \ref{fig:dminPlots} (left). %However, DM may have a drastically different local substructure that remains consistent with the Standard Halo Model on larger scales, and alternative models exist for the local properties of UDM.

We also consider an alternative UDM model, where it is assumed that UDM forms a local halo centered on the Sun~\cite{banerjee2020relaxion}. It is possible that DM forms gravitationally bounded objects, which could potentially form a halo bound to an external gravitational source, such as the Sun or Earth~\cite{banerjee2020relaxion}, while maintaining consistency with the Standard Halo Model on larger scales. In this scenario, the local DM halo could have a greater density and coherence time at Earth than in the Galactic halo model, leading to potentially stronger constraints on coupling strength from direct detection experiments. The halo's size and density would generally depend on the mass of the DM particles. Earth halos are well-motivated at particle energies $10^{-13}\,{\rm eV}\lesssim m_\text{DM}c^2\lesssim 10^{-7}$ eV~\cite{banerjee2020relaxion}, and several recent experiments have set constraints on the Earth halo model for DM~\cite{aiello2021constraints,savalle2021searching,oswald2021search,antypas2019scalar,aharony2021constraining}. A fiber-based detector is aptly suited to search for a local DM halo that is gravitationally bound to the Sun, as such Solar halos are well-motivated in the particle energy range $ 10^{-17}\,{\rm eV}\lesssim m_\text{DM}c^2\lesssim 10^{-13}$ eV~\cite{banerjee2020relaxion}. In this range, a Solar halo could have an energy density up to $\sim10^5$ times greater than that of a Galactic halo. The projected constraints from optical fibers, using a Solar halo model for DM, are plotted in Fig. \ref{fig:dminPlots} (right). To recalculate $d_{e,\text{min}}$ for a Solar DM halo, we re-scale $\rho_\text{DM}$ using Supplementary Figure 2 from Ref.~\cite{banerjee2020relaxion} and use $\tau_\text{DM} = 10^8 \text{s} \left(\frac{\rm 1 Hz}{f_\text{DM}}\right)^3$~\cite{banerjee2020searching}.

The parameter space for scalar UDM is plotted in Fig. \ref{fig:dminPlots} for both Galactic (left) and Solar (right) DM halos. The allowable parameter space is constrained by previous experiments, including both DM direct detection experiments and DM-independent equivalence principle (EP) tests. Experiments designed to test the equivalence principle~\cite{schlamminger2008test,berge2018microscope} set constraints on scalar interactions regardless of whether the scalar field composes DM. The strongest EP-test constraints in this mass range come from MICROSCOPE~\cite{berge2018microscope}, which has constrained $d_e\lesssim 10^{-4}$ for $m_\text{DM} \lesssim 10^{-13}$ eV/$c^2$. This constraint does not depend on the DM energy density $\rho_\text{DM}$, so it applies equally to both halo models in Fig. \ref{fig:dminPlots}. Also included are constraints from the DM direct detection experiments in Refs.~\cite{kennedy2020precision,filzinger2023improved,sherrill2023analysis}, which depend on the DM halo model considered. 

The potential of fiber-based scalar UDM detectors operating for $\tau_\text{exp} = 1$ year can be seen in Fig. \ref{fig:dminPlots}, where projections for $d_{e,\text{min}}$ are plotted for various experimental parameters. Fiber-based detectors operating at room temperature and $50$ mK, with and without polymer coatings, are considered to search for DM in the particle energy range $10^{-17} \text{ eV} \lesssim m_\text{DM}c^2 \lesssim 10^{-13}$ eV $\left(10^{-3}\text{ Hz}\lesssim f_\text{DM} \lesssim 10 \text{ Hz}\right)$. To enable room-temperature optical fibers to set more competitive constraints on $d_e$, an array of $N$ detectors can be implemented improving the sensitivity as $N^{-1/2}$ ~\cite{carney2021ultralight}. The expected constraints from an array of 10 independent detectors with bare fibers are included in green. %The fiber-based detectors proposed here can potentially probe new regions of the parameter space below $\sim 10^{-1}$ Hz.

\section{Summary and Outlook} \label{sec:conclusion}
In summary, we have investigated optical fibers as detectors for scalar ultralight dark matter. Dark matter-induced oscillations in the fine-structure constant $\alpha$ would produce a potentially measurable oscillation in fibers' optical path lengths that could be detected in a fiber-based interferometer. To estimate the detection capabilities of optical fibers, we propose an idea for a detector that could measure oscillations in the relative refractive indices of two different types of fibers that are differentially affected by scalar UDM. Accounting for various noise sources, we calculated the minimum detectable scalar UDM coupling strength considering two different UDM models (Galactic and Solar halos), finding that cryogenically cooled fiber-based detectors can achieve sensitivities to scalar UDM that exceed the current constraints at sub-Hz frequencies.

In addition to cryogenic cooling, competitive sensitivities can be achieved by using longer fibers or selecting (or fabricating) a better choice of optical fiber. The ideal fiber would likely have a thicker cladding and lower mechanical loss to reduce thermomechanical noise, and low optical loss to reduce shot noise and photothermal heating effects. Finally, the compactness of optical fibers facilitates array-based detection, where a network of detectors can be deployed to improve the overall sensitivity as well as to probe for local substructure in the UDM field.

Optical fibers are a mature technology with a multitude of applications in classical and quantum communication. Increasingly low tolerance for optical loss is driving various technological developments, making them a viable candidate for precision measurements beyond those relying on low photon number quantum optical effects. Creative ideas to harness such technology to study fundamental science are already emerging, such as those involving optical clock networks\cite{delva2017test, lisdat2016clock}. Along with using fibers as delay lines \cite{savalle2021searching} to look for scalar UDM or as a waveguide to search for axions \cite{yavuz2022generation}, our proposed interferometric scheme demonstrates that existing optical fiber technology can be repurposed to search for ultralight dark matter. Taken together, these developments present a promising avenue to harness this widely-used technology in the search for beyond the Standard Model physics.

%Finally, we acknowledge that cooling a kg-scale detector (potentially absorbing $\sim$mW optical power) to 10 mK is an ambitious undertaking that may require advances in cryogenic technology. However, to achieve competitive sensitivities while forgoing an aggressive cryogenics operation, one could implement an array of detectors, use longer fibers, or potentially find (or fabricate) a better choice of optical fiber. The ideal fiber would likely have a thicker cladding and lower mechanical loss to reduce thermomechanical noise, and low optical loss to reduce shot noise and photothermal heating in a cryogenic environment. Looking forward, we note that further research into the mechanical properties of optical fibers, particularly without polymer coatings and at cryogenic temperatures, would be beneficial in improving our estimates for thermomechanical noise. 

\section*{Acknowledgements}
We thank Yevgeny Stadnik, Pierre Dub{\'e}, Gregory Jasion, Dalziel Wilson, William Renninger, Matthew Konkol, William Beardell, Sean Nelan, Matthew Doty, Daniel Grin, and Dennis Prather for helpful discussions. This work is supported by the National Science Foundation Grants PHY-1912480, PHY-2047707, and the Office of the Under Secretary of Defense for Research and Engineering under award number FA9550-22-1-0323.

\appendix

\section*{Appendix}
This Appendix provides additional information on the proposed detection scheme, including a derivation of the signal and noise expressions, detail on the modulation of refractive indices by scalar UDM, and our estimation of the mechanical properties of optical fibers (necessary for estimating the thermomechanical noise floor). 

Appendix \ref{app:DMindex} provides a model for how we expect scalar UDM to modulate the fibers' refractive indices. Appendix \ref{phaseDifferenceAppendix} presents a model for the proposed detector and a derivation for the phase difference, including contributions from both scalar UDM and various noise sources.  Appendix \ref{app:noiseSources} contains extra information on the noise sources we have considered while estimating the sensitivity of a fiber-based detector. %Appendix \ref{app:fiberProperties} provides information on our estimates of various properties of optical fibers that are necessary to estimate thermomechanical noise. 
Appendix \ref{app:SignalAnalysis} provides an analysis for how the minimum detectable signal depends on the measurement time.

%-------------------------------------------------------------
\subsection{Definitions and model assumptions} \label{sec:definitionsAssumptions}
%-------------------------------------------------------------
\textit{Assumption 1:} We restrict our analysis to low frequencies, defined by $f_\text{DM}\ll {\tau_{0,i}}^{-1}$; i.e. this derivation holds when the DM Compton frequency is significantly below the inverse time delay of both fibers.

\textit{Assumption 2:} All fluctuations are small: 

$$\left\{\frac{\delta x}{x_0},\frac{\delta L_i}{L_{0,i}},\frac{\delta n_i}{n_{0,i}},\frac{\delta \omega_\text{L}}{\omega_{0,\text{L}}}\right\}\ll 1.$$

\textit{Definition of $\epsilon$-coefficients:} Consider a quantity $w(v)$ with steady-state value $w_0\equiv w(v_0)$ determined by the steady-state value of its parameter $v_0$.  Assuming small fluctuations ($\delta v \ll v_0$) in $v$ about $v_0$, the fractional fluctuations of $w$ can be expressed to first order as 
\begin{align*}
    \frac{\delta w}{w_0} \approx \epsilon_{wv}\frac{\delta v}{v_0}
\end{align*}
where 
\begin{equation}
\epsilon_{wv}\equiv \frac{v_0}{w_0}\frac{\partial w}{\partial v}\bigg|_{v=v_0}.
\end{equation}
Note that while $\epsilon_{wv}$ is dimensionless and resembles the sensitivity coefficients discussed in Ref.~\cite{kozlov2019comment}, the $\epsilon$-coefficients used in this work will not generally be independent in the choice of units. We use SI units throughout. We also define the quantities $\Delta \epsilon_{wv} \equiv \epsilon_{wv,\text{A}}-\epsilon_{wv,\text{B}}$. 

\begin{table}[ht]
	\centering
	\begin{tabular}{|l@{\hskip .75cm} l@{\hskip .75cm}|}
		\hline
		Effect 		& Expression\\
		\hline	
		DM-index coupling 	&	$\frac{\delta n_i}{n_{0,i}}=\epsilon_{nx,i}\frac{\delta x}{x_0}$\\
		DM-length coupling 	&	$\frac{\delta L_i}{L_{0,i}}=\epsilon_{Lx,i}\frac{\delta x}{x_0}$\\
		Mechanical vibrations	&	$\frac{\delta L_i}{L_{0,i}}=h_i^\text{acoustic}$\\
		Vibrations (strain-optic)     &   $\frac{\delta n_i}{n_{0,i}}=\epsilon_{nL,i}h_i^\text{acoustic}$\\
		Optical dispersion	& 	$\frac{\delta n_i}{n_{0,i}}=\epsilon_{n\omega_\text{L},i}\frac{\delta \omega_\text{L}}{\omega_{0,\text{L}}}$	\\	
		Fiber thermal noise	&	$\delta \phi_i=\delta \phi_i^\text{thermal}$ \\
		Shot noise (at detector)	&	$\delta \phi_i=\delta \phi_i^\text{shot}$ \\
		\hline				
	\end{tabular}
	\caption{Summary of fluctuations included in the the derivation of $\Delta\Phi_\text{A}$, due to both noise and DM. Each effect is expressed as a fluctuation in length, refractive index, or simply optical phase.}
	\label{tab:effects}
\end{table}

%-------------------------------------------------------------
%-------------------------------------------------------------
%\section{Refractive index effects} 
%-------------------------------------------------------------
\section{DM-coupling to refractive index} 
%\label{app:indexEffects}
\label{app:DMindex}
%-------------------------------------------------------------
The dependence of a dielectric medium's refractive index on fundamental constants such as $\alpha$ and $m_e$ can be estimated using a Lorentz model, which assumes that the medium is a collection of resonant atoms responding harmonically to an applied electric field. The refractive index depends on the electric susceptibility $n^2 = \chi + 1$, which accounts for multiple electron and phonon modes by considering the sum over all modes, $\chi = \sum_k \chi_k$. The susceptibility of each mode takes the following form off-resonance~\cite{braxmaier2001proposed}: 
\begin{equation}
    \chi_k = 4 \pi \hbar c \alpha\frac{N_k}{M_k} \frac{f_k}{{\omega_k}^2 - \omega^2}.
\end{equation}
Here, $N_k$ is the effective number density, $M_k$ is the reduced effective mass, $f_k$ is the oscillator strength, and $\omega_k$ is the resonance frequency. The number density is inversely proportional to the atomic/molecular volume [determined by the Bohr radius $a_\text{B} \propto \left(\alpha m_e\right)^{-1}$], such that $N_k \propto \left(\alpha m_e\right)^3$. For electron modes, $M_k=m_e$; for phonon modes, $M_k$ is independent of $\alpha$, $m_e$, and $\omega$~\cite{braxmaier2001proposed}. The oscillator strengths $f_k$ are also approximately independent of $\alpha$, $m_e$, and $\omega$~\cite{braxmaier2001proposed}. For both electron and phonon modes, it can be shown that
\begin{equation} \label{lorentzModel}
\chi_k \propto \left(1-\omega^2{\omega_k}^{-2}\right)^{-1}.
\end{equation}
Therefore, the dependence on fundamental constants $\alpha$ and $m_e$ is entirely accounted for by the resonance frequencies $\omega_k$. For phonon modes $\omega_k \propto {m_e}^{3/2}\alpha^2$; for electron modes $\omega_k \propto {m_e}\alpha^2$~\cite{braxmaier2001proposed}.

If the laser frequency is assumed to be in a range where it is reasonable to consider either only phonon modes or only electron modes (i.e. relationship between $\omega_k$ and the fundamental constants is independent of mode number $k$), Eq. \ref{lorentzModel} implies that 
\begin{equation} \label{generalDispersionFC}
\frac{x}{n}\frac{\partial n}{\partial x} = - \left(  \frac{x}{\omega_k}\frac{\partial \omega_k}{\partial x}\right) \left(\frac{\omega}{n}\frac{\partial n}{\partial \omega}\right)
\end{equation}
where $x\in \left\{\alpha,m_e\right\}$. Equation \ref{generalDispersionFC} allows one to calculate the response of a material's refractive index to variations of fundamental constants using the material's dispersion, which can be measured experimentally.

For variations in $\alpha$,
\begin{equation}
\frac{\alpha}{n}\frac{\partial n}{\partial \alpha} = - 2 \left(\frac{\omega}{n}\frac{\partial n}{\partial \omega}\right),
\end{equation}
regardless of the type of mode. However, the refractive index's dependence on $m_e$ depends on which modes dominate: 
\begin{equation}
\frac{m_e}{n}\frac{\partial n}{\partial m_e} =  \begin{cases}
-\left(\frac{\omega}{n}\frac{\partial n}{\partial \omega}\right) & \text{electron modes}\\
-\frac{3}{2}\left(\frac{\omega}{n}\frac{\partial n}{\partial \omega}\right) & \text{phonon modes}
\end{cases}
\end{equation}
Following Ref.~\cite{grote2019novel}, it is assumed that electron modes dominate the dispersion of fused silica, in which case 
\begin{align} 
&\epsilon_{n\alpha,\text{silica}} = -2\epsilon_{n\omega_\text{L},\text{silica}} \label{alphaIndex} \\
&\epsilon_{nm_e,\text{silica}} = -\epsilon_{n\omega_\text{L},\text{silica}}. \label{meIndex}
\end{align}
From the three-term Sellmeier equation for fused silica~\cite{malitson1965interspecimen}, $\epsilon_{n\omega_\text{L},\text{silica}} = 0.013$. Since the optical dispersion and, therefore, DM-coupling to the refractive index of air, are significantly less than that of silica, it is assumed that $\left\{\epsilon_{nx,\text{air}}, \epsilon_{n\omega_\text{L},\text{air}}\right\}=0$. 

The DM signal (Eq. \ref{appDMphase}) is proportional to $\epsilon_{nx,\text{silica}}+\epsilon_{n\omega_\text{L},\text{silica}}$. Evidently, from Eqs. \ref{alphaIndex} and \ref{meIndex}, the detector is only sensitive to DM-induced oscillations in $\alpha$, not $m_e$, since $\epsilon_{nm_e,\text{silica}}+\epsilon_{n\omega_\text{L},\text{silica}} = (-\epsilon_{n\omega_\text{L},\text{silica}}) +\epsilon_{n\omega_\text{L},\text{silica}}=0$. We note that a more detailed analysis (beyond the simple Lorentz model used here) of the refractive index of silica may affect this result.

%-------------------------------------------------------------
%-------------------------------------------------------------
\section{Derivation of output phase difference $\Delta\Phi_\text{A}$} \label{phaseDifferenceAppendix}
\begin{figure}[ht]
	\centering
	\includegraphics[width=0.98 \columnwidth,clip, trim= 0in 1.5in 0in 1.4in]{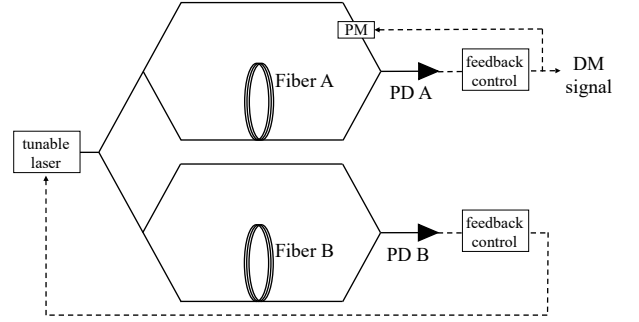}
	\caption{\textbf{Detector model with FDL-stabilized laser:} The purpose of this detector geometry is to eliminate the effects of laser phase noise relative to the geometry proposed in Fig. \ref{fig:balancedSchematic}. Fiber B serves as a delay line in an interferometer to stabilize a tunable laser. Using feedback from the measurement at Photodetector B (PD B), the laser frequency is tuned such that $\Delta \Phi_\text{B}$ remains constant. The phase difference $\Delta \Phi_\text{A}$ would then contain the DM signal. The phase difference $\Delta \Phi_\text{A}$ depends linearly on the optical power when the recombining beams are kept near quadrature bias ($\Delta \Phi_\text{A}$ mod $2\pi$ = $\pi/2$). In the presence of noise and long term instabilities, quadrature bias can be maintained by including a phase modulator (PM) that is driven by feedback from Photodetector A (PD A). The DM signal is then present in the feedback signal supplied to the phase modulator. }
	\label{fig:schematic}
\end{figure}

%-------------------------------------------------------------
%-------------------------------------------------------------
In this section we derive an expression for the output phase difference $\Delta\Phi_\text{A}$ for the detector displayed in Fig. \ref{fig:schematic} in terms of fluctuations due to both noise and the dark matter signal.

This analysis is general to DM-modulation of both electron mass $m_e$ and the fine-structure constant $\alpha$; fundamental constants will be generally represented by the variable $x\in \left\{m_e,\alpha\right\}$. While it was shown in Appendix \ref{app:DMindex} that for our specific choice of detector configuration and the model for refractive index, the  detector is not sensitive to oscillation in $m_e$, a different material choice and/or configuration might give access to the $d_{m_e}$ coupling. The analysis presented here can be applied to other material-based interferometer schemes with minimal modifications.

For simplicity, we introduced the detector as a simple optical-length-balanced interferometer in the main text (see Fig. \ref{fig:balancedSchematic}). In order to cancel the detrimental effects of a variety of technical noise sources, we now assume a more complex geometry as shown in Fig. \ref{fig:schematic}. Our detailed analysis in this and the next section will reveal that several technical noise sources can be suppressed via common mode rejection in this, more complex, configuration. In essence, the detector geometry displayed in Fig. \ref{fig:schematic} has a sensitivity that is consistent with a balanced interferometer (represented in Fig. \ref{fig:balancedSchematic}), while being immune to laser frequency noise. We note that the quantitative results presented in Figs. \ref{fig:phaseNoisePlot} and \ref{fig:dminPlots} come from the derivations presented here, which assume the experimental geometry of Fig. \ref{fig:schematic}.

\subsection{Fractional fluctuations of fiber refractive index and length}
Fiber $i$'s refractive index $n_i \left(x,\omega_\text{L}\right)$ depends on the fundamental constant $x$, the laser frequency $\omega_\text{L}$ due to optical dispersion, and vibration-induced strains $h_i^\text{acoustic}$ via the strain-optic effect:
\begin{equation} \label{fractionalIndexChange}
    \frac{\delta n_i}{n_{0,i}} = \epsilon_{nx,i}\frac{\delta x}{x_0} + \epsilon_{n\omega_\text{L},i}\frac{\delta\omega_\text{L}}{\omega_{0,\text{L}}}+\epsilon_{nL,i}h_i^\text{acoustic}.
\end{equation}
Additionally, the fiber lengths depend on fundamental constant $x$, as the length of a solid is proportional to the Bohr radius, which will be affected by fluctuations in $m_e$ or $\alpha$ ($a_\text{B}\propto \left(m_e\alpha)^{-1}\right)$~\cite{stadnik2015searching, arvanitaki2016sound}.
Since $\epsilon_{Lx,i}=-1$ for both fibers, the differential scheme described in Fig. \ref{fig:balancedSchematic} is insensitive to this strain signal.

\subsection{Optical phase at fiber output}
\begin{figure}[ht]
	{\centering
	\includegraphics[width=0.98 \columnwidth]{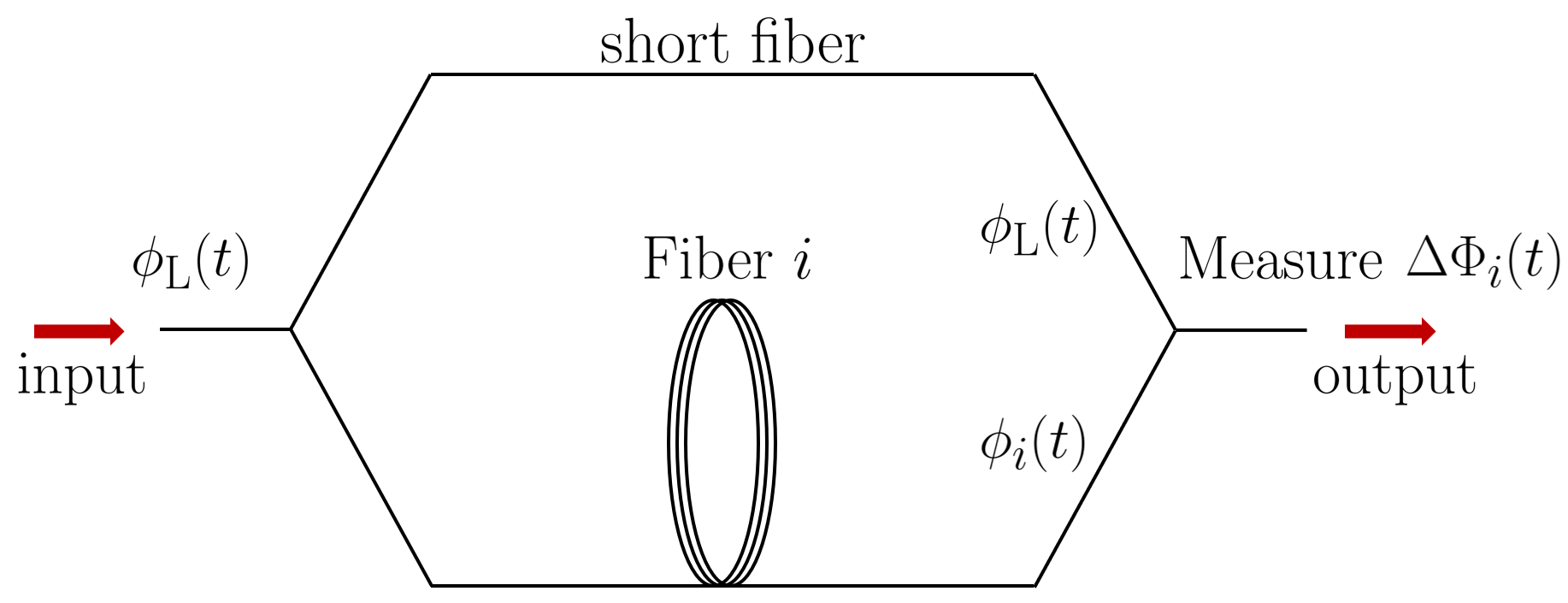}
	\caption{\textbf{Illustration of fiber interferometer, with relevant phases marked:} The inferred output phase difference $\Delta \Phi_i(t) = \phi_\text{L}(t) - \phi_i(t) +\delta\phi_i^\text{shot}$ is determined by measuring the output optical power.}
	\label{fig:fiberPhase} 
}
\end{figure}

Consider a simple fiber interferometer as depicted in Fig. \ref{fig:fiberPhase}. If the phase of light entering a fiber is $\phi_\text{L}(t)$, then the phase of light exiting the fiber will be equal to the input phase at an earlier time $\phi_i(t) =\phi_\text{L}(t-\tau_i(t))$, where $\tau_i$ is the time it takes for an optical wavefront to traverse the entire length of the fiber. In the low frequency limit (Assumption 1) the time delay is simply related to the fiber's instantaneous optical path length $\tau_i(t) =n_i(t)L_i(t)/c$, and the output phase is
$$\phi_i(t) =  \phi_\text{L}(t) - \omega_\text{L}(t) n_i(t)L_i(t)/c.$$ 
Including small fluctuations of the optical frequency, fiber index, and fiber length, the output optical phase can be written as
\begin{align*}
\phi_i \approx \phi_\text{L}- \omega_{0,\text{L}}\tau_{0,i}\left(1 + \frac{\delta n_i}{n_{0,i}} + \frac{\delta L_i}{L_{0,i}}+ \frac{\delta \omega_\text{L}}{\omega_{0,\text{L}}}\right).
\end{align*}
We have omitted from our notation the explicit time dependence of each variable $(t)$, since the analysis considers the quasi-static (low frequency) regime.

Accounting for both noise and DM-induced effects (Table \ref{tab:effects}), the optical phase at the output of Fiber $i$ is
\begin{widetext}
\begin{equation}	\label{fiberPhase}
   \phi_i = \underbrace{\phi_\text{L}}_{\substack{\text{input}\\\text{laser}\\\text{phase}}}
- \omega_{0,\text{L}}\tau_{0,i}\left(
\underbrace{1}_{\text{DC term}}
+\underbrace{\epsilon_{nx,i}\frac{\delta x}{x_0}}_{\substack{\text{DM-index}\\\text{coupling}}}
+\underbrace{\epsilon_{n\omega_\text{L},i}\frac{\delta \omega_\text{L}}{\omega_{0,\text{L}}}}_{\substack{\text{optical}\\\text{dispersion}}}
+\underbrace{\epsilon_{Lx,i}\frac{\delta x}{x_0}}_{\substack{\text{DM-length}\\\text{coupling}}}
+\underbrace{\left(1+\epsilon_{nL,i}\right)h_i^\text{acoustic}}_{\substack{\text{mechanical}\\\text{vibrations}}}
+\underbrace{\frac{\delta \omega_\text{L}}{\omega_{0,\text{L}}}}_{\substack{\text{laser frequency}\\\text{fluctuations}}}
\right) 
+\underbrace{\delta \phi_i^\text{thermal}}_{\substack{\text{fiber thermal}\\\text{fluctuations}}}
\end{equation}
\end{widetext}

\subsection{Laser stabilized to Fiber B}
For this analysis, the schematic in Fig. \ref{fig:schematic} is assumed (see Appendix \ref{app:laserNoise}), where Fiber B acts as a delay line for stabilizing the laser. The inferred phase difference at Photodetector B is $\Delta\Phi_\text{B}=\phi_\text{L} - \phi_\text{B} + \delta \phi_\text{B}^\text{shot}$, where $\phi_\text{B}^\text{shot}$ is the apparent phase imparted by shot noise. Stabilization can be achieved by tuning the laser's frequency using feedback based on the optical power measured by Photodetector B. For example, the laser's frequency can be tuned such that $\Delta \Phi_\text{B}$ mod $2\pi$ = $\pi/2$, maintaining quadrature bias between the recombining beams before Photodetector B. At quadrature bias, where the output optical power depends linearly on the phase difference $\Phi_\text{B}$, it can be shown (using Eq. \ref{fiberPhase}) that the laser frequency fluctuations will be
\begin{widetext}
\begin{equation}	\label{laserFluctuations}
\frac{\delta \omega_\text{L}}{\omega_{0,\text{L}}}=\left(1+\epsilon_{n\omega_\text{L},\text{B}}\right)^{-1}\left(\frac{\frac{\pi}{2}+\delta \phi_\text{B}^\text{thermal} - \delta \phi_\text{B}^\text{shot}}{\omega_{0,\text{L}}\tau_{0,\text{B}}}-\left(1+\epsilon_{nL,\text{B}}\right)h_\text{B}^\text{acoustic}-\left(\epsilon_{nx,\text{B}}-1\right)\frac{\delta x}{x_0}\right).
\end{equation}
\end{widetext}

%-------------------------------------------------------------
\subsection{Detector output phase difference $\Delta\Phi_\text{A}$}
%-------------------------------------------------------------
The experimental observable in this experiment is the differential phase $\Delta\Phi_\text{A}$ between the output of Fiber A and the interferometer's short arm, which is inferred from the measured optical power at Photodetector A. Here it is assumed that quadrature bias is maintained between the recombining beams before Photodetector A, which can be accomplished with a phase modulator that is driven by feedback from the power measured at Photodetector A. The DM signal would then be present in the feedback signal supplied to the modulator.%mention Geraci paper 

The phase difference $\Delta \Phi_\text{A}$ can be calculated using Eqs. \ref{fiberPhase} and \ref{laserFluctuations}. Ignoring zero-frequency terms, $\Delta\Phi_\text{A}$ is the sum of  individual contributions from DM effects and each noise source
$$\Delta\Phi_\text{A}= \Delta\Phi_\text{A}^\text{DM} + \Delta\Phi_\text{A}^\text{thermal} + \Delta\Phi_\text{A}^\text{acoustic} + \Delta\Phi_\text{A}^\text{shot}.$$
Noting that $\epsilon_{nx}\sim \epsilon_{n\omega_\text{L}}\approx 10^{-2}$ and ignoring higher-order terms, it can be shown that
\begin{align}
    \Delta \Phi_\text{A}^\text{DM} &\approx\omega_{0,\text{L}}\tau_{0} \left(\Delta \epsilon_{nx}+\Delta \epsilon_{n\omega_\text{L}}\right)\frac{\delta x}{x_0} \label{appDMphase}	\\
    \Delta \Phi_\text{A}^\text{acoustic} &\approx \!\begin{multlined}[t][1cm]
        \omega_{0,\text{L}}\tau_{0}\left[\left(1+\epsilon_{nL,\text{A}}\right)h_\text{A}^\text{acoustic}  \right. \\
        -\left.\left(1+\Delta \epsilon_{n\omega_\text{L}}+\epsilon_{nL,\text{B}}\right)h_\text{B}^\text{acoustic}\right]	
    \end{multlined} \label{acousticPhaseEquation}\\
    \Delta \Phi_\text{A}^\text{thermal} &\approx - \delta \phi_\text{A}^\text{thermal}+\delta \phi_\text{B}^\text{thermal} 	\\
    \Delta \Phi_\text{A}^\text{shot} &\approx\delta \phi_\text{A}^\text{shot}-\delta \phi_\text{B}^\text{shot}, 	
\end{align}

where we have assumed both fibers to have equal optical path lengths $\tau_{0,i}\equiv \tau_0$ for simplicity.

%-------------------------------------------------------------
%-------------------------------------------------------------
\section{Noise analysis} \label{app:noiseSources}
%-------------------------------------------------------------
%-------------------------------------------------------------
In this section we discuss the potential sources of noise that would act to obscure the DM-induced phase difference. The effect of each noise source is quantified by a one-sided phase noise power spectral density (PSD) $S_{\Delta \Phi,\text{A}}(f)$.

%-------------------------------------------------------------
\subsection{Laser noise} 
\label{app:laserNoise}
%-------------------------------------------------------------
Interferometric experiments are subject to noise from the driving laser, due to random fluctuations in the optical phase/frequency and intensity. Laser phase/frequency noise is eliminated in a perfectly balanced interferometer. However, when each arm is composed of different types of fiber with different levels of optical dispersion, an interferometer cannot remain balanced for all optical frequencies and laser frequency noise is non-negligible. For the sensitivity of a balanced interferometer to reach the thermal noise floor at $50$ mK, laser frequency noise needs to be limited to roughly $\sim 10^{-1} \left(\text{Hz}/f\right)^{1/2}$ Hz$\cdot$Hz$^{-1/2}$; this would require an ultrastable laser~\cite{wu20160}.

Alternatively, laser frequency noise can be mitigated by stabilizing the frequency-tunable laser to one of the fibers, which acts as a delay line in a Mach-Zehnder interferometer. As a result, the laser's frequency noise PSD takes the form of the fiber delay line's (FDL)\footnote{Ultra-stable lasers have been demonstrated using FDL-stabilization. For example, a linewidth of 200 mHz has been achieved with a 5 km FDL at room temperature~\cite{huang2019all}.} phase noise PSD:\footnote{This relationship holds for low frequencies, as the bandwidth of the control loop is limited by the time delay in the FDL~\cite{dong2015subhertz}: $f\ll {\tau_0}^{-1}\approx 20$ kHz.} $S_{\delta\nu,\text{L}} = \left(2 \pi \tau_0\right)^{-2} S_{\delta \phi,\text{FDL}}$~\cite{dong2015subhertz}. This procedure effectively eliminates laser frequency noise from the analysis, replacing it with phase noise from Fiber B (which would have been present to the same level in a balanced interferometer). 

For the detector depicted in Fig. \ref{fig:schematic}, the ``observable" is the phase difference $\Delta \Phi_\text{A}$ between the recombining optical paths before Photodetector A, which would still be affected by DM-induced modulation of the fiber refractive indices. At low frequencies $\left(f_\text{DM}\ll {\tau_0}^{-1}\right)$, this is simply $\Delta \Phi_\text{A} = 2 \pi n_\text{A}L_\text{A} / \lambda_{\text{L}}$ (prior to receiving corrections from the phase modulator). Because the laser is stabilized to Fiber B, the laser wavelength is proportional to the optical path length of Fiber B: $\lambda_\text{L}\propto n_\text{B}L_\text{B}$. Therefore, the detector's function is to compare the optical path lengths of Fibers A and B, as $\Delta \Phi_\text{A}$ is proportional to the ratio $n_\text{A}L_\text{A}/n_\text{B}L_\text{B}$. The balanced interferometer displayed in Fig. \ref{fig:balancedSchematic} is still a reasonable conceptual model for the detector, and it can be seen from the derivation for $\Delta \Phi_\text{A}$ in Appendix \ref{phaseDifferenceAppendix} the magnitude of the DM signal $\Delta \Phi^\text{DM}$ is unchanged in Eq. \ref{dmPhaseSignal}.

%As shown in Appendix \ref{phaseDifferenceAppendix}, the noise sources that typically plague unbalanced fiber interferometers, such as acoustic noise, can still be reduced through common mode suppression in the setup from Fig. \ref{fig:schematic}. Therefore, the detector geometry displayed in Fig. \ref{fig:schematic} has a sensitivity that is consistent with a balanced interferometer that is immune to laser frequency noise. For simplicity, we present the detector as a simple optical-length-balanced interferometer in the main text, and note that the quantitative results presented in Figs. \ref{fig:phaseNoisePlot} and \ref{fig:dminPlots} come from the derivations in Appendix \ref{phaseDifferenceAppendix}, which assume the experimental geometry of Fig. \ref{fig:schematic}.

In addition to laser frequency noise, random fluctuations of the laser's optical intensity would limit the sensitivity of a detector like that depicted in Fig. \ref{fig:schematic}. Experiments for measuring thermal noise in fibers have successfully minimized laser intensity noise to below the thermal noise floor with methods such as heterodyne detection~\cite{dong2016observation} or with the use of differential amplifiers~\cite{bartolo2012thermal}. Assuming these methods are implemented, laser intensity noise is neglected in this analysis.

We note that the resulting appearance of the detector in Fig. \ref{fig:schematic} is similar to that of the DAMNED experiment~\cite{savalle2021searching} for scalar UDM, where an ultrastable laser drives an unbalanced fiber interferometer. There are, however, some key differences. While the detector we propose is sensitive to DM-induced refractive index modulation at sub-Hz frequencies, the DAMNED experiment is primarily sensitive to the DM-induced strain at kHz frequencies where their optical cavity has acoustic resonances. Schematically, the only major difference is the use of an optical fiber for stabilizing the laser instead of an ultrastable optical cavity. However, this distinction is important for suppressing acoustic noise at low frequencies (explained in Section \ref{sec:vibrations}) to reach the fibers' thermal noise floor. By achieving a thermally-limited measurement scheme, the sensitivity of the detector can then be enhanced through cryogenic cooling. In fact, operating at low temperatures will likely be necessary to set novel constraints on scalar UDM at sub-Hz frequencies, as demonstrated by the results in Fig. \ref{fig:dminPlots}. 

%-------------------------------------------------------------
\subsection{Photon shot noise and optical power}
%-------------------------------------------------------------
The phase difference $\Delta \Phi$ between the arms of an interferometer is inferred by measuring the output optical power $P$ after the junction where the arms recombine. Assuming that the recombining beams have equal optical power and are maintained at quadrature bias,\footnote{At quadrature bias, the average phase difference between recombining beams at an interferometer's output is $\left<\Delta \Phi\right>$ mod $2\pi$ = $\pi/2$.} the PSDs of the phase difference and output optical power fluctuations $\Delta P$ are related by $S_{\Delta \Phi}=S_{\Delta P}/\left<P\right>^2$, where $\left<P\right>$ is the average output optical power.  

Small fluctuations in the measured optical power due to shot noise are indistinguishable from small phase fluctuations in an interferometer, thereby limiting the sensitivity of our setup. %\footnote{Optical power fluctuations due to photon shot noise should be distinguished from laser intensity noise. The effects of photon shot noise are not reduced by the same methods that would reduce the effects of laser intensity noise, such as heterodyne detection~\cite{dong2016observation} or balanced photodetection~\cite{bartolo2012thermal}.}
The one-sided PSD of the fluctuations in measured optical power due to shot noise is~\cite{clerk2010introduction}
\begin{equation} \label{photonshotnoise}
S_{\Delta P}^\text{shot} = \frac{2 h c}{\lambda_{0,\text{L}}} \left<P\right>.
\end{equation}
Assuming that the optical power imminent on each photodetector is equal, 
\begin{equation}
\left<P\right> = \frac{P_\text{L}}{2 + 10^{\gamma_\text{B}L_\text{B}}+10^{\gamma_\text{A}L_\text{A}}},
\end{equation}
where $\gamma_i$ is the optical loss of Fiber $i$ and $P_\text{L}$ is the optical power of the laser. The total contribution of shot noise to the differential phase noise of the detector is thus
\begin{equation}
S_{\Delta \Phi,\text{A}}^\text{shot} = \frac{4 h c }{\lambda_{0,\text{L}}P_\text{L}}\left(2 + 10^{\gamma_\text{B}L_\text{B}}+10^{\gamma_\text{A}L_\text{A}}\right),
\end{equation}
which is simply double the shot noise at each detector.

The effects of shot noise can be reduced by increasing the optical power. Figure \ref{fig:phaseNoisePlot} shows that a detector with bare fibers, operating at $T=50$ mK, achieves a noise floor (combined thermomechanical and acoustic noise) of $\gtrsim 10^{-7} {\rm rad / \sqrt{\rm Hz}}$. Reducing the total shot noise in this experiment to the subdominant level of $\sqrt{S_{\Delta \Phi,\text{A}}^\text{shot}}\lesssim 10^{-7} {\rm rad / \sqrt{\rm Hz}}$ requires an average optical power on each photodetector of $\left<P\right>\gtrsim 50\, \mu$W. 

Recent advances in hollow core fiber technology have yielded hollow core fibers with a propagation loss of $0.17$ dB/km ($\gamma = 1.7\times 10^{-5} \, {\rm m^{-1}}$) at $\lambda_{0,\text{L}}=1550$ nm~\cite{jasion20220}, reaching the level of low-loss solid core fibers ($0.18$ dB/km for SMF28~\cite{smf28}), and we use these numbers for our shot noise estimates. In Fig. \ref{fig:phaseNoisePlot}, the total shot noise is plotted for a detector with a $10$ km solid core and $\sim15$ km hollow core fiber, assuming a laser power of $P_\text{L}=1$ mW. To account for optical loss, the optical power being diverted into the solid and hollow core fibers is taken to be $0.25$ mW and $0.42$ mW, respectively, and the power reaching each photodetector, $\left<P\right>= 0.16$ mW.

%-------------------------------------------------------------
\subsection{Thermal noise in fibers} 
\label{sectionThermalNoise}
%-------------------------------------------------------------
The detector's noise floor in the $10^{-3}-10^1$ Hz frequency range will likely be dominated by thermal noise in the optical fibers, which will induce optical phase fluctuations $\delta \phi$ at the fiber outputs. Thermal noise in fibers includes contributions from two effects, which are referred to as ``thermomechanical" noise $S_{\delta\phi}^\text{TM}$ (discussed in the main text) and ``thermoconductive" noise $S_{\delta\phi}^\text{TC}$~\cite{duan2012general}. Thermoconductive noise results from spontaneous local temperature fluctuations within an optical fiber~\cite{kittel1988temperature,glenn1989noise} that affect the fiber's length and refractive index via thermal expansion and the thermo-optic effect, respectively~\cite{duan2012general}. We have estimated $S_{\delta\phi}^\text{TC}$ in solid core fibers using the parameters (for SMF28) and Eq. 1 from Ref.~\cite{dong2016observation}. Thermoconductive noise in hollow core fibers will likely be lower than that of solid core fibers at lower frequencies $\left( \lesssim 1 \text{ kHz}\right)$, especially in HCFs that are evacuated~\cite{michaud2022fundamental}. 

Assuming the thermal noise for each fiber is uncorrelated,% and therefore add in quadrature,
\begin{equation} \label{fiberThermalPSDs}
    S_{\Delta\Phi,\text{A}}^\text{thermal} = \sum_i S_{\delta\phi,i}^\text{TM}+S_{\delta\phi,i}^\text{TC}.
\end{equation}

Thermoconductive noise is approximately frequency-independent at lower frequencies $\left( \lesssim 1 \text{ kHz}\right)$, and scales with temperature and fiber length as $S_{\delta\phi}^\text{TC}\propto T^2 L$~\cite{wanser1992fundamental}. Thermomechanical noise, scaling as $S_{\delta\phi}^\text{TM}\propto T L/f$, is expected to dominate the experiment given lower frequencies and temperatures considered here. This can be seen in the thermal noise curves in Fig. \ref{fig:phaseNoisePlot}, which inherit the $1/f$ frequency scaling from thermomechanical noise, except for the 300 K bare case, where thermoconductive noise dominates above $\sim 10^{-2}$ Hz. 

\begin{table*}[t]
	\centering
	\begin{tabular}{| l@{\hskip .5cm} l@{\hskip .5cm} c@{\hskip .5cm}  c@{\hskip .5cm}|}
		\hline
		\textbf{Quantity}	&	\textbf{Variable [Units]}			& \textbf{Solid Core} & \textbf{Hollow Core} 		\\
		\hline	
		refractive index &$n_{0}$ [-]		& $1.47$~\cite{smf28} &  $1$  \\
		optical loss    &   $\gamma$ [m$^{-1}$] &   $1.8 \times 10^{-5}$~\cite{smf28}    &  $2.8 \times 10^{-5}$~\cite{jasion2020hollow}      \\
		optical dispersion  &   $\epsilon_{n\omega_\text{L}}$ [-]   & $0.013$~\cite{malitson1965interspecimen}  & 0  \\
		stress-optic effect     & $\epsilon_{nL}$ [-]   &    $-0.17$~\cite{butter1978fiber}  &   0   \\
		mechanical loss tangent &$\xi$ [-]		& $10^{-2}$ 		\cite{beadle2001measurement,dong2016observation}		& $10^{-2}$* \\
		diameter of coating &$d_\text{coat}$ [$\mu$m]	& $242$~\cite{smf28}	& $302$*  \\
		diameter of cladding &$d_\text{clad}$	[$\mu$m]	& $125$~\cite{smf28}	&  $185$~\cite{shi2021thinly}  \\
		diameter of photonic crystal region &$d_\text{pc}$	[$\mu$m]	& $0$	& $70$~\cite{shi2021thinly}  \\
		total cross-sectional area&$A$ [$\mu$m$^2$]		& $4.6 \times 10^{4}$	& $6.8 \times 10^{4}$ \\		
		Young's modulus &$E$ [GPa]	& $20$ &   $25$ \\
		\hline				
	\end{tabular}
	\caption{Parameters used for room temperature optical fibers with coatings. The solid core fiber is assumed to be SMF28~\cite{smf28}, and the hollow core fiber is assumed to be nested antiresonant nodeless fiber (NANF)~\cite{jasion2020hollow}. We also discuss how the parameters are affected by removing the fiber coatings and operating at cryogenic temperatures. *The loss tangent and coating thickness for hollow core fiber is assumed equal to those of SMF28.}
	\label{tab:fiberparameters}
\end{table*} 

Key parameters for evaluating fiber thermomechanical noise using Eq. \ref{thermomechanicalNoise} are summarized in Table \ref{tab:fiberparameters}. While the physical properties $\xi$ and $E$ have not been  measured specifically for SMF28 fibers, the estimates found in Table \ref{tab:fiberparameters} have been shown to reliably predict thermomechanical noise in SMF28 fibers~\cite{dong2016observation} via Eq. \ref{thermomechanicalNoise}. Below we provide details for the evaluation of $E$, and the choice of $\xi$ for various scenarios considered in this work.

\emph{Young's modulus evaluation:}
The Young's modulus $\left(E\right)$ is approximated for a fiber by a weighted average of the Young's moduli of each material over the cross-sectional area~\cite{beadle2001measurement}:
\begin{equation} \label{youngs}
E \approx \frac{A_\text{coat}E_\text{coat}+A_\text{clad}E_\text{clad}}{A}.
\end{equation}
The values $E_\text{clad}=66.1$ GPa for a silica fiber and $E_\text{coat}=3.3$ GPa for the polymer coating (assuming acrylate) are used, based on measurements at kHz frequencies~\cite{beadle2001measurement}. The Young's moduli are assumed to be frequency-independent. The values for $E$ in Table \ref{tab:fiberparameters} are calculated from Eq. \ref{youngs} with the following cross-sectional areas 
\begin{align*}
A&\approx \frac{\pi}{4}\left({d_\text{coat}}^2-{d_\text{pc}}^2\right) \\ A_\text{coat}&=\frac{\pi}{4}\left({d_\text{coat}}^2-{d_\text{clad}}^2\right)  \\
A_\text{clad} &\approx \frac{\pi}{4}\left({d_\text{clad}}^2-{d_\text{pc}}^2\right),
\end{align*}
which approximate the hollow and photonic crystal regions to be empty space. 

\emph{Bare fiber parameters:}
While fused silica optical fibers with a polymer-coating have a loss tangent $\xi\approx 10^{-2}$~\cite{beadle2001measurement,dong2016observation}, bulk fused silica can potentially achieve $\xi\approx 10^{-8}$ at room temperature~\cite{schroeter2007mechanical}. In addition, mechanical dissipation in the violin modes (transverse oscillations) of bare fibers at room temperature is typically around $\xi \approx 10^{-7}-10^{-6}$~\cite{gretarsson1999dissipation, penn2001high, heptonstall2010investigation}. Noting that mechanical loss in a spooled optical fiber will likely differ from that of bulk silica or violin modes in silica suspensions under tension, a value of $\xi= 10^{-6}$ is assumed for bare fibers at room temperature. By setting $d_\text{coat} = d_\text{clad}$ the fibers' cross-sectional areas and effective Young's moduli are recalculated, with all other parameters in Table \ref{tab:fiberparameters} unchanged. 

\emph{Low-temperature parameters:}
Measurements of the temperature dependence of the Young's modulus of fused silica~\cite{mcskimin1953measurement} and the refractive index of fused silica fibers~\cite{reid1998temperature} suggest that $n_0$ and $E_\text{clad}$ can be assumed constant with respect to temperature without appreciably affecting the estimates in Fig. \ref{fig:phaseNoisePlot}. %However, the mechanical loss tangent $\xi$ of fused silica at kHz-MHz frequencies has been shown to exhibit a strong temperature dependence, with a maximum of $\xi\sim 10^{-3}$ at around 30-50 K~\cite{schroeter2007mechanical,arcizet2009cryogenic} before beginning to decrease with lowering temperature beyond that, although it is unclear how well this translates to fibers. In this work, it is considered to use thinner of fiber coatings as a way to reduce the loss tangent, approaching that of fused silica. 
We assume $\xi= 10^{-3}$ for cryogenic ($50$ mK) bare fibers, based on measurements in Ref.~\cite{behunin2017engineering}.

At low temperatures ($\sim 200$ K), the acrylate coatings transition from a rubbery state to a stiffer, glassy state with a Young's modulus $E_\text{coat} \approx 40$ GPa~\cite{zhu2019thermal}, and the effective Young's modulus has been adjusted for coated fibers at $50$ mK accordingly with Eq. \ref{youngs}. However, due to the absence of data on mechanical dissipation in optical fibers at low temperatures, we assume $\xi=10^{-2}$ for coated fibers at all temperatures.

%-------------------------------------------------------------
\subsection{Acoustic noise in fibers}  
\label{sec:vibrations}
%-------------------------------------------------------------
Mechanical vibrations alter the optical path lengths of fibers, resulting in phase noise. Here we estimate the amplitude of mechanical vibrations, which we will generically refer to as ``acoustic noise," and discuss techniques to potentially reduce the resulting phase noise to a subdominant level. 

To estimate the effects of mechanical vibrations, we use the USGS New High Noise Model (NHNM) for seismic noise, which can be found in Table 4 in Ref.~\cite{peterson1993observations}. The NHNM provides an estimate of seismic acceleration noise $S_{aa}^\text{seismic}$ for a hypothetical, relatively noisy location on Earth. For the frequency range considered in this work ($\sim 10^{-3}-10^1$ Hz), the NHNM roughly predicts an upper bound of
\begin{equation} \label{seismicUpperBound}
S_{aa}^\text{seismic}\lesssim 10^{-9} \left(\frac{\rm m}{\rm s^2}\right)^2 {\rm Hz}^{-1},
\end{equation}
%where both the horizontal and vertical components of the acceleration are approximately equal in magnitude~\cite{peterson1993observations,newell1997ultra}. The NHNM in Ref.~\cite{peterson1993observations} extends from $10^{-5}$ Hz to $10$ Hz, 
which we extrapolate to frequencies above $10$ Hz by assuming white acceleration noise $\sim 10^{-9} \left(\frac{\rm m}{\rm s^2}\right)^2 {\rm Hz}^{-1}$. We will assume this seismic noise accounts for all of the vibrations the fibers will experience: $S_{aa}^\text{acoustic}\approx S_{aa}^\text{seismic}$

Vibration-induced strains in optical fibers can be suppressed by using low vibration sensitivity fiber spools, a technology designed to reduce frequency noise in fiber delay line (FDL)-stabilized lasers~\cite{li2011low}. This technique has been shown to produce ultrastable lasers at room temperature, reaching the thermomechanical noise floor in km-scale fibers at sub-Hz frequencies~\cite{huang2019all}. Such vibration-insensitive fiber spools have achieved sensitivities $\Gamma_a\equiv h^\text{acoustic}/a^\text{acoustic} \lesssim 10^{-11} \left(\rm m/s^2\right)^{-1}$ to both horizontal and vertical accelerations~\cite{huang2019vibration}. Thus, we expect acoustic noise to induce a strain noise in each fiber on the order of
\begin{equation} \label{acousticStrainNoise}
S_{hh}^\text{acoustic}=\left|\Gamma_a\right|^2 S_{aa}^\text{acoustic}\lesssim 10^{-31} {\rm \, Hz}^{-1}.
\end{equation}

%-------------------------------------------------------------
%\emph{Strain-optic effect:} 
%\label{app:strainOptic}
%-------------------------------------------------------------
In solid core fiber, this will lead to a change in the refractive index via the strain-optic effect
\begin{equation}
\frac{\delta n}{n_0} = \epsilon_{nL}h^\text{acoustic}.
\end{equation}
From the analysis in Ref.~\cite{butter1978fiber}, which considers a linear strain along a single mode fiber, it can be shown that $\epsilon_{nL,\text{solid}}\approx -0.17.$ 

Accounting for the strain-optic effect $\epsilon_{nL,i}$ and differences in optical dispersion $\Delta\epsilon_{n\omega_\text{L}}$, it can be shown that the output phase difference due to strains $h_i^\text{acoustic}$ in both fibers is (see Appendix \ref{phaseDifferenceAppendix})
\begin{multline}
    \Delta \Phi_\text{A}^\text{acoustic} \approx
        \omega_{0,\text{L}}\tau_{0}\left[\left(1+\epsilon_{nL,\text{A}}\right)h_\text{A}^\text{acoustic}  \right. \\
        -\left.\left(1+\Delta \epsilon_{n\omega_\text{L}}+\epsilon_{nL,\text{B}}\right)h_\text{B}^\text{acoustic}\right],	
\end{multline}
%suggesting that acoustic noise can be suppressed by a factor $\left(\Delta \epsilon_{n\omega_\text{L}}+\Delta \epsilon_{nL}\right)\approx 10^{-1}$ (see Appendix \ref{app:indexEffects}) in an experiment where 

The strain-optic effect has a larger impact on the acoustic phase noise than the optical dispersion effect $\left(\big|\Delta \epsilon_{nL}\big|>\big|\Delta \epsilon_{n\omega_\text{L}}\big|\right)$.

If the individual fiber strains are correlated and of equal magnitude $h_\text{A}^\text{acoustic}=h_\text{B}^\text{acoustic}$,
\begin{equation} \label{acousticPhaseNoise}
S_{\Delta \Phi,\text{A}}^\text{acoustic} = \left|\omega_{0,\text{L}}\tau_{0}\left(\Delta \epsilon_{n\omega_\text{L}}+\Delta \epsilon_{nL}\right) \right|^2 S_{hh}^\text{acoustic}.
\end{equation}

This can be accomplished by co-winding the fibers on a spool or cylinder, a technique that has been used to observe the thermal noise floor in fiber-based interferometers~\cite{bartolo2012thermal,dong2016observation}. By co-winding both fibers on a single spool with $\Gamma_a \lesssim 10^{-11} \left(\rm m/s^2\right)^{-1}$, acoustic noise at the level predicted by the NHNM can be reduced to a level comparable to that of thermal noise in this experiment, as can be seen in Fig. \ref{fig:phaseNoisePlot}. If desired, further suppression of vibration-induced phase noise can also be achieved using the feedforward method~\cite{thorpe2010measurement,leibrandt2013cavity}.

%-------------------------------------------------------------
%-------------------------------------------------------------
 \section{Minimum detectable signal vs measurement time}
\label{app:SignalAnalysis}
%-------------------------------------------------------------
%-------------------------------------------------------------
The purpose of this appendix is to motivate Eq. \ref{phiminEquation}, which describes how the minimum detectable signal $\Delta \Phi_\text{min}^\text{DM}$ scales with the measurement time $\tau_\text{exp}$. There are two separate regimes for dark matter detection, the ``stochastic" $\left(\tau_\text{exp} \lesssim \tau_\text{DM}\right)$ and ``deterministic" $\left(\tau_\text{exp} \gg \tau_\text{DM}\right)$ regimes \cite{centers2021stochastic}. As discussed in Ref. \cite{budker2014proposal}, $\Delta \Phi_\text{min}^\text{DM}\propto {\tau_\text{exp}}^{-1/2}$ for a coherent signal and $\Delta \Phi_\text{min}^\text{DM}\propto {\tau_\text{exp}}^{-1/4}$ for measurement durations greatly exceeding the DM coherence time. Here we provide a complementary analysis using a discrete Fourier transform (DFT) formalism and join both regimes with a single expression for the DM-induced phase difference.

We consider the case where the measurement is contaminated by noise $\Delta\Phi^\text{noise}$.  Performing a measurement of duration $\tau_\text{exp}$ with a sampling rate of $f_s$, a periodogram $S_{\Delta\Phi,k}^\text{noise}$ can be obtained from the DFT $\Delta \tilde{\Phi}_k^\text{noise}$ as $S_{\Delta\Phi,k}^\text{noise} = 2|\Delta \tilde{\Phi}_k^\text{noise}|^2/ N f_s$, where $N\equiv f_s \tau_\text{exp}$. Here, the subscript $k$ refers to the bin number. Assuming Gaussian noise, the periodogram's standard deviation can be approximated by its mean~\cite{DSPBook}, which for a broadband noise source can be estimated simply by sampling the PSD: $\left<S_{\Delta\Phi,k}^\text{noise}\right>\approx S_{\Delta \Phi}^\text{noise}(kf_s/N)$. This expression is valid in the large $N$ limit.

Here we define the minimum detectable signal strength $\Delta \Phi_\text{min}^\text{DM}$ as the signal amplitude required for the signal power to equal the variance in the noise floor, i.e. 
\begin{equation} \label{SNR1}
    S_{\Delta\Phi,k}^\text{DM} = S_{\Delta \Phi}^\text{noise}(kf_s/N).
\end{equation}
In the following sections we explore the relationship between $\Delta \Phi_\text{min}^\text{DM}$ and the expected noise floor $S_{\Delta \Phi}^\text{noise}(kf_s/N)$ for the stochastic and deterministic regimes.

\begin{figure}[b]
	\centering
	\includegraphics[width=0.98 \columnwidth,clip, trim= 0in 0in 0in 0in]{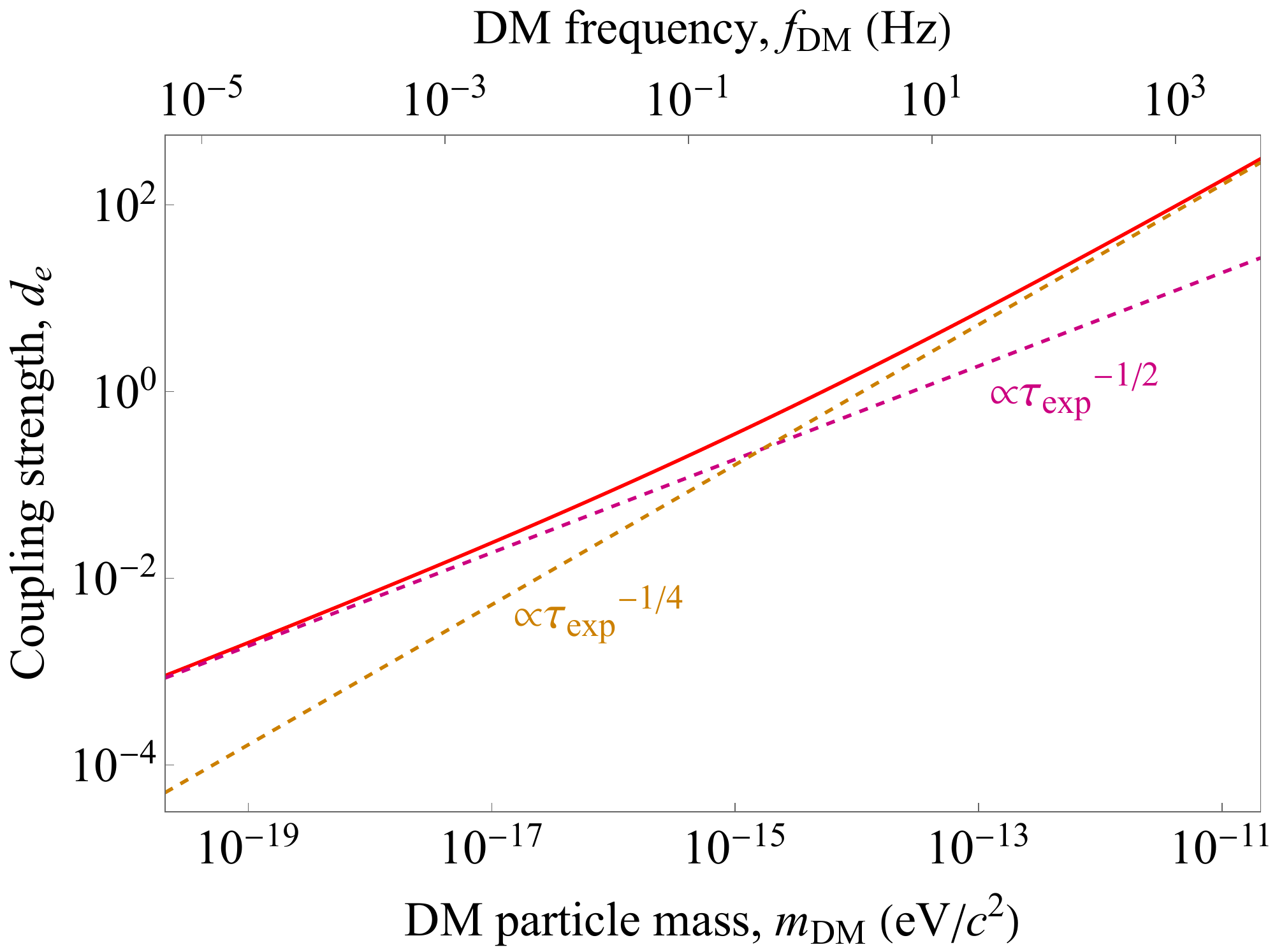}
	\caption{\textbf{Dependence of $d_{e,\text{min}}$ on $\tau_\text{exp}$:} We approximate the minimum detectable signal strength's dependence on the total measurement time with Eq. \ref{phiminEquation} (Eq. \ref{eq:combinedphimin}), which accounts for both measurement regimes: $\tau_\text{exp}\lesssim \tau_\text{DM}$ (where $\Delta \Phi_\text{min}^\text{DM}\propto {\tau_\text{exp}}^{-1/2}$) and $\tau_\text{exp}\gg \tau_\text{DM}$ (where $\Delta \Phi_\text{min}^\text{DM}\propto {\tau_\text{exp}}^{-1/4}$). Here, $d_{e,\text{min}}$ is calculated using Eq. \ref{stocasticPhiMin} (dashed pink line), Eq. \ref{deterministicPhiMin} (dashed yellow line), and Eq. \ref{eq:combinedphimin} (red curve), assuming a noise floor limited purely by the thermomechanical noise of room temperature, coated fibers. It can be seen that the combined expression from Eq. \ref{eq:combinedphimin} converges to the $\propto {\tau_\text{exp}}^{-1/2}$ and $\propto {\tau_\text{exp}}^{-1/4}$ limits at low and high frequencies, respectively. }
	\label{fig:scaling}
\end{figure}

\subsection{Minimum detectable signal for $\tau_\text{exp} \lesssim \tau_\text{DM}$}
For measurement times less than the DM coherence time the signal is approximately coherent $\Delta\Phi^\text{DM} \approx  \Delta\Phi^\text{DM}_0 \cos\left(2 \pi f_\text{DM} t + \theta_\text{DM}\right)$. For a coherent signal, the signal power is entirely contained within a single frequency bin,
\begin{equation}
    S_{\Delta\Phi,k}^\text{DM} = \frac{1}{2}\left(\Delta \Phi_0^\text{DM}\right)^2 \tau_\text{exp}.
\end{equation}
From Eq. \ref{SNR1}, the minimum detectable signal would be $\Delta \Phi_\text{min}^\text{DM} \approx \sqrt{2S_{\Delta \Phi}^\text{noise}}\tau_\text{exp}^{-1/2}$.

This analysis so far assumed a deterministic signal amplitude $\Delta \Phi_0^\text{DM}$. However, due to interference of UDM field components of differing frequency and phase, the field's amplitude fluctuates over timescales roughly equal to $\tau_\text{DM}$. The field amplitude within a chunk of time $\tau_\text{DM}$, $\varphi_0$, is a stochastic quantity with a Rayleigh distribution, where $\sim 63\%$ of all realizations are less than $\varphi_\text{rms}$, resulting in a reduced signal~\cite{centers2021stochastic}. Therefore, a DM signal in the stochastic regime would need an amplitude $\sim3$ times larger than an otherwise identical signal with deterministic amplitude to have the same probability of detection~\cite{centers2021stochastic}. Accounting for the stochastic nature of the UDM field, the minimum detectable signal is 
\begin{equation}
\label{stocasticPhiMin}
 \Delta \Phi_\text{min}^\text{DM} \approx 3\sqrt{2S_{\Delta \Phi}^\text{noise}} \tau_\text{exp}^{-1/2}
\end{equation}

\subsection{Minimum detectable signal for $\tau_\text{exp} \gg \tau_\text{DM}$}
For measurement times exceeding the DM coherence time, the signal power would be spread across $\sim \tau_\text{exp}/\tau_\text{DM}$ frequency bins. However, the measurement could be split into $M=\tau_\text{exp}/\tau_\text{DM}$ shorter measurements of duration $\tau_\text{DM}$. This procedure is known as Bartlett's method (described in Ref. \cite{DSPBook}), where the total periodogram would be the average of the $M$ short-time periodograms, resulting in an $M$-fold reduction in the variance in the noise floor $S_{\Delta \Phi}^\text{noise}\rightarrow S_{\Delta \Phi}^\text{noise}/\sqrt{M}$. Each individual data segment is duration $\tau_\text{DM}$, so $S_{\Delta\Phi,k}^\text{DM} \approx \left(\Delta \Phi_0^\text{DM}\right)^2 \tau_\text{DM}/2$. The minimum detectable signal amplitude is then
\begin{equation}\label{deterministicPhiMin}
    \Delta \Phi_\text{min}^\text{DM} \approx \sqrt{2S_{\Delta \Phi}^\text{noise}}(\tau_\text{DM}\tau_\text{exp})^{-1/4}
\end{equation}

\subsection{Combined expression for  $\Delta \Phi_\text{min}^\text{DM}$}
For measurement times less than the DM coherence time, the minimum detectable signal scales as $\propto {\tau_\text{exp}}^{-1/2}$ (Eq. \ref{stocasticPhiMin}). For measurement times greatly exceeding the DM coherence time, the minimum detectable signal scales as $\propto {\tau_\text{exp}}^{-1/4}$ (Eq. \ref{deterministicPhiMin}). This relationship between $\Delta \Phi_\text{min}^\text{DM}$ and $\tau_\text{exp}$ is a broken power law, which we simplify with a ``smoothly joined broken power law"~\cite{brokenPowerLaw} by simply adding contributions from both regimes as
\begin{equation}
    \Delta \Phi_\text{min}^\text{DM} \approx \sqrt{2S_{\Delta \Phi}^\text{noise}}\left[ 3\tau_\text{exp}^{-1/2} + (\tau_\text{DM}\tau_\text{exp})^{-1/4}\right].
    \label{eq:combinedphimin}
\end{equation}
This function reproduces the behavior of Eqs. \ref{stocasticPhiMin} and \ref{deterministicPhiMin} in the limits $\tau_\text{exp} \ll \tau_\text{DM}$ and $\tau_\text{exp} \gg \tau_\text{DM}$, respectively, while resulting in a larger $\Delta \Phi_\text{min}^\text{DM}$ in the $\tau_\text{exp} \approx \tau_\text{DM}$ regime, as shown in Fig. \ref{fig:scaling}. This effect is desirable, as there would be some carry over of stochastic effects in this regime.

%\nocite{*}
\bibliography{references}

\begin{thebibliography}{77}%
\makeatletter
\providecommand \@ifxundefined [1]{%
 \@ifx{#1\undefined}
}%
\providecommand \@ifnum [1]{%
 \ifnum #1\expandafter \@firstoftwo
 \else \expandafter \@secondoftwo
 \fi
}%
\providecommand \@ifx [1]{%
 \ifx #1\expandafter \@firstoftwo
 \else \expandafter \@secondoftwo
 \fi
}%
\providecommand \natexlab [1]{#1}%
\providecommand \enquote  [1]{``#1''}%
\providecommand \bibnamefont  [1]{#1}%
\providecommand \bibfnamefont [1]{#1}%
\providecommand \citenamefont [1]{#1}%
\providecommand \href@noop [0]{\@secondoftwo}%
\providecommand \href [0]{\begingroup \@sanitize@url \@href}%
\providecommand \@href[1]{\@@startlink{#1}\@@href}%
\providecommand \@@href[1]{\endgroup#1\@@endlink}%
\providecommand \@sanitize@url [0]{\catcode `\\12\catcode `\$12\catcode
  `\&12\catcode `\#12\catcode `\^12\catcode `\_12\catcode `\%12\relax}%
\providecommand \@@startlink[1]{}%
\providecommand \@@endlink[0]{}%
\providecommand \url  [0]{\begingroup\@sanitize@url \@url }%
\providecommand \@url [1]{\endgroup\@href {#1}{\urlprefix }}%
\providecommand \urlprefix  [0]{URL }%
\providecommand \Eprint [0]{\href }%
\providecommand \doibase [0]{http://dx.doi.org/}%
\providecommand \selectlanguage [0]{\@gobble}%
\providecommand \bibinfo  [0]{\@secondoftwo}%
\providecommand \bibfield  [0]{\@secondoftwo}%
\providecommand \translation [1]{[#1]}%
\providecommand \BibitemOpen [0]{}%
\providecommand \bibitemStop [0]{}%
\providecommand \bibitemNoStop [0]{.\EOS\space}%
\providecommand \EOS [0]{\spacefactor3000\relax}%
\providecommand \BibitemShut  [1]{\csname bibitem#1\endcsname}%
\let\auto@bib@innerbib\@empty
%</preamble>
\bibitem [{\citenamefont {Group}\ \emph {et~al.}(2020)\citenamefont {Group},
  \citenamefont {Zyla}, \citenamefont {Barnett}, \citenamefont {Beringer},
  \citenamefont {Dahl}, \citenamefont {Dwyer}, \citenamefont {Groom},
  \citenamefont {Lin}, \citenamefont {Lugovsky}, \citenamefont {Pianori} \emph
  {et~al.}}]{particle2020review}%
  \BibitemOpen
  \bibfield  {author} {\bibinfo {author} {\bibfnamefont {P.~D.}\ \bibnamefont
  {Group}}, \bibinfo {author} {\bibfnamefont {P.}~\bibnamefont {Zyla}},
  \bibinfo {author} {\bibfnamefont {R.}~\bibnamefont {Barnett}}, \bibinfo
  {author} {\bibfnamefont {J.}~\bibnamefont {Beringer}}, \bibinfo {author}
  {\bibfnamefont {O.}~\bibnamefont {Dahl}}, \bibinfo {author} {\bibfnamefont
  {D.}~\bibnamefont {Dwyer}}, \bibinfo {author} {\bibfnamefont
  {D.}~\bibnamefont {Groom}}, \bibinfo {author} {\bibfnamefont {C.-J.}\
  \bibnamefont {Lin}}, \bibinfo {author} {\bibfnamefont {K.}~\bibnamefont
  {Lugovsky}}, \bibinfo {author} {\bibfnamefont {E.}~\bibnamefont {Pianori}},
  \emph {et~al.},\ }\href@noop {} {\bibfield  {journal} {\bibinfo  {journal}
  {Progress of Theoretical and Experimental Physics}\ }\textbf {\bibinfo
  {volume} {2020}},\ \bibinfo {pages} {083C01} (\bibinfo {year}
  {2020})}\BibitemShut {NoStop}%
\bibitem [{\citenamefont {Battaglieri}\ \emph {et~al.}(2017)\citenamefont
  {Battaglieri}, \citenamefont {Belloni}, \citenamefont {Chou}, \citenamefont
  {Cushman}, \citenamefont {Echenard}, \citenamefont {Essig}, \citenamefont
  {Estrada}, \citenamefont {Feng}, \citenamefont {Flaugher}, \citenamefont
  {Fox} \emph {et~al.}}]{battaglieri2017us}%
  \BibitemOpen
  \bibfield  {author} {\bibinfo {author} {\bibfnamefont {M.}~\bibnamefont
  {Battaglieri}}, \bibinfo {author} {\bibfnamefont {A.}~\bibnamefont
  {Belloni}}, \bibinfo {author} {\bibfnamefont {A.}~\bibnamefont {Chou}},
  \bibinfo {author} {\bibfnamefont {P.}~\bibnamefont {Cushman}}, \bibinfo
  {author} {\bibfnamefont {B.}~\bibnamefont {Echenard}}, \bibinfo {author}
  {\bibfnamefont {R.}~\bibnamefont {Essig}}, \bibinfo {author} {\bibfnamefont
  {J.}~\bibnamefont {Estrada}}, \bibinfo {author} {\bibfnamefont {J.~L.}\
  \bibnamefont {Feng}}, \bibinfo {author} {\bibfnamefont {B.}~\bibnamefont
  {Flaugher}}, \bibinfo {author} {\bibfnamefont {P.~J.}\ \bibnamefont {Fox}},
  \emph {et~al.},\ }\href@noop {} {\bibfield  {journal} {\bibinfo  {journal}
  {arXiv preprint arXiv:1707.04591}\ } (\bibinfo {year} {2017})}\BibitemShut
  {NoStop}%
\bibitem [{\citenamefont {Antypas}\ \emph {et~al.}(2022)\citenamefont
  {Antypas}, \citenamefont {Banerjee}, \citenamefont {Bartram}, \citenamefont
  {Baryakhtar}, \citenamefont {Betz}, \citenamefont {Bollinger}, \citenamefont
  {Boutan}, \citenamefont {Bowring}, \citenamefont {Budker}, \citenamefont
  {Carney} \emph {et~al.}}]{antypas2022new}%
  \BibitemOpen
  \bibfield  {author} {\bibinfo {author} {\bibfnamefont {D.}~\bibnamefont
  {Antypas}}, \bibinfo {author} {\bibfnamefont {A.}~\bibnamefont {Banerjee}},
  \bibinfo {author} {\bibfnamefont {C.}~\bibnamefont {Bartram}}, \bibinfo
  {author} {\bibfnamefont {M.}~\bibnamefont {Baryakhtar}}, \bibinfo {author}
  {\bibfnamefont {J.}~\bibnamefont {Betz}}, \bibinfo {author} {\bibfnamefont
  {J.}~\bibnamefont {Bollinger}}, \bibinfo {author} {\bibfnamefont
  {C.}~\bibnamefont {Boutan}}, \bibinfo {author} {\bibfnamefont
  {D.}~\bibnamefont {Bowring}}, \bibinfo {author} {\bibfnamefont
  {D.}~\bibnamefont {Budker}}, \bibinfo {author} {\bibfnamefont
  {D.}~\bibnamefont {Carney}},  \emph {et~al.},\ }\href@noop {} {\bibfield
  {journal} {\bibinfo  {journal} {arXiv preprint arXiv:2203.14915}\ } (\bibinfo
  {year} {2022})}\BibitemShut {NoStop}%
\bibitem [{\citenamefont {Damour}\ and\ \citenamefont
  {Polyakov}(1994)}]{damour1994string}%
  \BibitemOpen
  \bibfield  {author} {\bibinfo {author} {\bibfnamefont {T.}~\bibnamefont
  {Damour}}\ and\ \bibinfo {author} {\bibfnamefont {A.~M.}\ \bibnamefont
  {Polyakov}},\ }\href@noop {} {\bibfield  {journal} {\bibinfo  {journal}
  {Nuclear Physics B}\ }\textbf {\bibinfo {volume} {423}},\ \bibinfo {pages}
  {532} (\bibinfo {year} {1994})}\BibitemShut {NoStop}%
\bibitem [{\citenamefont {Van~Tilburg}\ \emph {et~al.}(2015)\citenamefont
  {Van~Tilburg}, \citenamefont {Leefer}, \citenamefont {Bougas},\ and\
  \citenamefont {Budker}}]{van2015search}%
  \BibitemOpen
  \bibfield  {author} {\bibinfo {author} {\bibfnamefont {K.}~\bibnamefont
  {Van~Tilburg}}, \bibinfo {author} {\bibfnamefont {N.}~\bibnamefont {Leefer}},
  \bibinfo {author} {\bibfnamefont {L.}~\bibnamefont {Bougas}}, \ and\ \bibinfo
  {author} {\bibfnamefont {D.}~\bibnamefont {Budker}},\ }\href@noop {}
  {\bibfield  {journal} {\bibinfo  {journal} {Physical review letters}\
  }\textbf {\bibinfo {volume} {115}},\ \bibinfo {pages} {011802} (\bibinfo
  {year} {2015})}\BibitemShut {NoStop}%
\bibitem [{\citenamefont {Hees}\ \emph {et~al.}(2016)\citenamefont {Hees},
  \citenamefont {Gu{\'e}na}, \citenamefont {Abgrall}, \citenamefont {Bize},\
  and\ \citenamefont {Wolf}}]{hees2016searching}%
  \BibitemOpen
  \bibfield  {author} {\bibinfo {author} {\bibfnamefont {A.}~\bibnamefont
  {Hees}}, \bibinfo {author} {\bibfnamefont {J.}~\bibnamefont {Gu{\'e}na}},
  \bibinfo {author} {\bibfnamefont {M.}~\bibnamefont {Abgrall}}, \bibinfo
  {author} {\bibfnamefont {S.}~\bibnamefont {Bize}}, \ and\ \bibinfo {author}
  {\bibfnamefont {P.}~\bibnamefont {Wolf}},\ }\href@noop {} {\bibfield
  {journal} {\bibinfo  {journal} {Physical review letters}\ }\textbf {\bibinfo
  {volume} {117}},\ \bibinfo {pages} {061301} (\bibinfo {year}
  {2016})}\BibitemShut {NoStop}%
\bibitem [{\citenamefont {Beloy}\ and\ \citenamefont
  {Bodine}(2021)}]{beloy2021frequency}%
  \BibitemOpen
  \bibfield  {author} {\bibinfo {author} {\bibfnamefont {K.}~\bibnamefont
  {Beloy}}\ and\ \bibinfo {author} {\bibfnamefont {M.~I.}\ \bibnamefont
  {Bodine}},\ }\href@noop {} {\bibfield  {journal} {\bibinfo  {journal}
  {Nature}\ }\textbf {\bibinfo {volume} {591}} (\bibinfo {year}
  {2021})}\BibitemShut {NoStop}%
\bibitem [{\citenamefont {Filzinger}\ \emph {et~al.}(2023)\citenamefont
  {Filzinger}, \citenamefont {D{\"o}rscher}, \citenamefont {Lange},
  \citenamefont {Klose}, \citenamefont {Steinel}, \citenamefont {Benkler},
  \citenamefont {Peik}, \citenamefont {Lisdat},\ and\ \citenamefont
  {Huntemann}}]{filzinger2023improved}%
  \BibitemOpen
  \bibfield  {author} {\bibinfo {author} {\bibfnamefont {M.}~\bibnamefont
  {Filzinger}}, \bibinfo {author} {\bibfnamefont {S.}~\bibnamefont
  {D{\"o}rscher}}, \bibinfo {author} {\bibfnamefont {R.}~\bibnamefont {Lange}},
  \bibinfo {author} {\bibfnamefont {J.}~\bibnamefont {Klose}}, \bibinfo
  {author} {\bibfnamefont {M.}~\bibnamefont {Steinel}}, \bibinfo {author}
  {\bibfnamefont {E.}~\bibnamefont {Benkler}}, \bibinfo {author} {\bibfnamefont
  {E.}~\bibnamefont {Peik}}, \bibinfo {author} {\bibfnamefont {C.}~\bibnamefont
  {Lisdat}}, \ and\ \bibinfo {author} {\bibfnamefont {N.}~\bibnamefont
  {Huntemann}},\ }\href@noop {} {\bibfield  {journal} {\bibinfo  {journal}
  {arXiv preprint arXiv:2301.03433}\ } (\bibinfo {year} {2023})}\BibitemShut
  {NoStop}%
\bibitem [{\citenamefont {Sherrill}\ \emph {et~al.}(2023)\citenamefont
  {Sherrill}, \citenamefont {Parsons}, \citenamefont {Baynham}, \citenamefont
  {Bowden}, \citenamefont {Curtis}, \citenamefont {Hendricks}, \citenamefont
  {Hill}, \citenamefont {Hobson}, \citenamefont {Margolis}, \citenamefont
  {Robertson} \emph {et~al.}}]{sherrill2023analysis}%
  \BibitemOpen
  \bibfield  {author} {\bibinfo {author} {\bibfnamefont {N.}~\bibnamefont
  {Sherrill}}, \bibinfo {author} {\bibfnamefont {A.~O.}\ \bibnamefont
  {Parsons}}, \bibinfo {author} {\bibfnamefont {C.~F.}\ \bibnamefont
  {Baynham}}, \bibinfo {author} {\bibfnamefont {W.}~\bibnamefont {Bowden}},
  \bibinfo {author} {\bibfnamefont {E.~A.}\ \bibnamefont {Curtis}}, \bibinfo
  {author} {\bibfnamefont {R.}~\bibnamefont {Hendricks}}, \bibinfo {author}
  {\bibfnamefont {I.~R.}\ \bibnamefont {Hill}}, \bibinfo {author}
  {\bibfnamefont {R.}~\bibnamefont {Hobson}}, \bibinfo {author} {\bibfnamefont
  {H.~S.}\ \bibnamefont {Margolis}}, \bibinfo {author} {\bibfnamefont {B.~I.}\
  \bibnamefont {Robertson}},  \emph {et~al.},\ }\href@noop {} {\bibfield
  {journal} {\bibinfo  {journal} {arXiv preprint arXiv:2302.04565}\ } (\bibinfo
  {year} {2023})}\BibitemShut {NoStop}%
\bibitem [{\citenamefont {Campbell}\ \emph {et~al.}(2021)\citenamefont
  {Campbell}, \citenamefont {McAllister}, \citenamefont {Goryachev},
  \citenamefont {Ivanov},\ and\ \citenamefont {Tobar}}]{campbell2021searching}%
  \BibitemOpen
  \bibfield  {author} {\bibinfo {author} {\bibfnamefont {W.~M.}\ \bibnamefont
  {Campbell}}, \bibinfo {author} {\bibfnamefont {B.~T.}\ \bibnamefont
  {McAllister}}, \bibinfo {author} {\bibfnamefont {M.}~\bibnamefont
  {Goryachev}}, \bibinfo {author} {\bibfnamefont {E.~N.}\ \bibnamefont
  {Ivanov}}, \ and\ \bibinfo {author} {\bibfnamefont {M.~E.}\ \bibnamefont
  {Tobar}},\ }\href@noop {} {\bibfield  {journal} {\bibinfo  {journal}
  {Physical Review Letters}\ }\textbf {\bibinfo {volume} {126}},\ \bibinfo
  {pages} {071301} (\bibinfo {year} {2021})}\BibitemShut {NoStop}%
\bibitem [{\citenamefont {Geraci}\ \emph {et~al.}(2019)\citenamefont {Geraci},
  \citenamefont {Bradley}, \citenamefont {Gao}, \citenamefont {Weinstein},\
  and\ \citenamefont {Derevianko}}]{geraci2019searching}%
  \BibitemOpen
  \bibfield  {author} {\bibinfo {author} {\bibfnamefont {A.~A.}\ \bibnamefont
  {Geraci}}, \bibinfo {author} {\bibfnamefont {C.}~\bibnamefont {Bradley}},
  \bibinfo {author} {\bibfnamefont {D.}~\bibnamefont {Gao}}, \bibinfo {author}
  {\bibfnamefont {J.}~\bibnamefont {Weinstein}}, \ and\ \bibinfo {author}
  {\bibfnamefont {A.}~\bibnamefont {Derevianko}},\ }\href@noop {} {\bibfield
  {journal} {\bibinfo  {journal} {Physical review letters}\ }\textbf {\bibinfo
  {volume} {123}},\ \bibinfo {pages} {031304} (\bibinfo {year}
  {2019})}\BibitemShut {NoStop}%
\bibitem [{\citenamefont {Kennedy}\ \emph {et~al.}(2020)\citenamefont
  {Kennedy}, \citenamefont {Oelker}, \citenamefont {Robinson}, \citenamefont
  {Bothwell}, \citenamefont {Kedar}, \citenamefont {Milner}, \citenamefont
  {Marti}, \citenamefont {Derevianko},\ and\ \citenamefont
  {Ye}}]{kennedy2020precision}%
  \BibitemOpen
  \bibfield  {author} {\bibinfo {author} {\bibfnamefont {C.~J.}\ \bibnamefont
  {Kennedy}}, \bibinfo {author} {\bibfnamefont {E.}~\bibnamefont {Oelker}},
  \bibinfo {author} {\bibfnamefont {J.~M.}\ \bibnamefont {Robinson}}, \bibinfo
  {author} {\bibfnamefont {T.}~\bibnamefont {Bothwell}}, \bibinfo {author}
  {\bibfnamefont {D.}~\bibnamefont {Kedar}}, \bibinfo {author} {\bibfnamefont
  {W.~R.}\ \bibnamefont {Milner}}, \bibinfo {author} {\bibfnamefont {G.~E.}\
  \bibnamefont {Marti}}, \bibinfo {author} {\bibfnamefont {A.}~\bibnamefont
  {Derevianko}}, \ and\ \bibinfo {author} {\bibfnamefont {J.}~\bibnamefont
  {Ye}},\ }\href@noop {} {\bibfield  {journal} {\bibinfo  {journal} {Physical
  Review Letters}\ }\textbf {\bibinfo {volume} {125}},\ \bibinfo {pages}
  {201302} (\bibinfo {year} {2020})}\BibitemShut {NoStop}%
\bibitem [{\citenamefont {Savalle}\ \emph {et~al.}(2021)\citenamefont
  {Savalle}, \citenamefont {Hees}, \citenamefont {Frank}, \citenamefont
  {Cantin}, \citenamefont {Pottie}, \citenamefont {Roberts}, \citenamefont
  {Cros}, \citenamefont {McAllister},\ and\ \citenamefont
  {Wolf}}]{savalle2021searching}%
  \BibitemOpen
  \bibfield  {author} {\bibinfo {author} {\bibfnamefont {E.}~\bibnamefont
  {Savalle}}, \bibinfo {author} {\bibfnamefont {A.}~\bibnamefont {Hees}},
  \bibinfo {author} {\bibfnamefont {F.}~\bibnamefont {Frank}}, \bibinfo
  {author} {\bibfnamefont {E.}~\bibnamefont {Cantin}}, \bibinfo {author}
  {\bibfnamefont {P.-E.}\ \bibnamefont {Pottie}}, \bibinfo {author}
  {\bibfnamefont {B.~M.}\ \bibnamefont {Roberts}}, \bibinfo {author}
  {\bibfnamefont {L.}~\bibnamefont {Cros}}, \bibinfo {author} {\bibfnamefont
  {B.~T.}\ \bibnamefont {McAllister}}, \ and\ \bibinfo {author} {\bibfnamefont
  {P.}~\bibnamefont {Wolf}},\ }\href@noop {} {\bibfield  {journal} {\bibinfo
  {journal} {Physical Review Letters}\ }\textbf {\bibinfo {volume} {126}},\
  \bibinfo {pages} {051301} (\bibinfo {year} {2021})}\BibitemShut {NoStop}%
\bibitem [{\citenamefont {Arvanitaki}\ \emph {et~al.}(2015)\citenamefont
  {Arvanitaki}, \citenamefont {Huang},\ and\ \citenamefont
  {Van~Tilburg}}]{arvanitaki2015searching}%
  \BibitemOpen
  \bibfield  {author} {\bibinfo {author} {\bibfnamefont {A.}~\bibnamefont
  {Arvanitaki}}, \bibinfo {author} {\bibfnamefont {J.}~\bibnamefont {Huang}}, \
  and\ \bibinfo {author} {\bibfnamefont {K.}~\bibnamefont {Van~Tilburg}},\
  }\href@noop {} {\bibfield  {journal} {\bibinfo  {journal} {Physical Review
  D}\ }\textbf {\bibinfo {volume} {91}},\ \bibinfo {pages} {015015} (\bibinfo
  {year} {2015})}\BibitemShut {NoStop}%
\bibitem [{\citenamefont {Manley}\ \emph {et~al.}(2020)\citenamefont {Manley},
  \citenamefont {Wilson}, \citenamefont {Stump}, \citenamefont {Grin},\ and\
  \citenamefont {Singh}}]{manley2020searching}%
  \BibitemOpen
  \bibfield  {author} {\bibinfo {author} {\bibfnamefont {J.}~\bibnamefont
  {Manley}}, \bibinfo {author} {\bibfnamefont {D.~J.}\ \bibnamefont {Wilson}},
  \bibinfo {author} {\bibfnamefont {R.}~\bibnamefont {Stump}}, \bibinfo
  {author} {\bibfnamefont {D.}~\bibnamefont {Grin}}, \ and\ \bibinfo {author}
  {\bibfnamefont {S.}~\bibnamefont {Singh}},\ }\href@noop {} {\bibfield
  {journal} {\bibinfo  {journal} {Physical review letters}\ }\textbf {\bibinfo
  {volume} {124}},\ \bibinfo {pages} {151301} (\bibinfo {year}
  {2020})}\BibitemShut {NoStop}%
\bibitem [{\citenamefont {Vermeulen}\ \emph {et~al.}(2021)\citenamefont
  {Vermeulen}, \citenamefont {Relton}, \citenamefont {Grote}, \citenamefont
  {Raymond}, \citenamefont {Affeldt}, \citenamefont {Bergamin}, \citenamefont
  {Bisht}, \citenamefont {Brinkmann}, \citenamefont {Danzmann}, \citenamefont
  {Doravari} \emph {et~al.}}]{vermeulen2021direct}%
  \BibitemOpen
  \bibfield  {author} {\bibinfo {author} {\bibfnamefont {S.~M.}\ \bibnamefont
  {Vermeulen}}, \bibinfo {author} {\bibfnamefont {P.}~\bibnamefont {Relton}},
  \bibinfo {author} {\bibfnamefont {H.}~\bibnamefont {Grote}}, \bibinfo
  {author} {\bibfnamefont {V.}~\bibnamefont {Raymond}}, \bibinfo {author}
  {\bibfnamefont {C.}~\bibnamefont {Affeldt}}, \bibinfo {author} {\bibfnamefont
  {F.}~\bibnamefont {Bergamin}}, \bibinfo {author} {\bibfnamefont
  {A.}~\bibnamefont {Bisht}}, \bibinfo {author} {\bibfnamefont
  {M.}~\bibnamefont {Brinkmann}}, \bibinfo {author} {\bibfnamefont
  {K.}~\bibnamefont {Danzmann}}, \bibinfo {author} {\bibfnamefont
  {S.}~\bibnamefont {Doravari}},  \emph {et~al.},\ }\href@noop {} {\bibfield
  {journal} {\bibinfo  {journal} {Nature}\ }\textbf {\bibinfo {volume} {600}},\
  \bibinfo {pages} {424} (\bibinfo {year} {2021})}\BibitemShut {NoStop}%
\bibitem [{\citenamefont {Aiello}\ \emph {et~al.}(2021)\citenamefont {Aiello},
  \citenamefont {Richardson}, \citenamefont {Vermeulen}, \citenamefont {Grote},
  \citenamefont {Hogan}, \citenamefont {Kwon},\ and\ \citenamefont
  {Stoughton}}]{aiello2021constraints}%
  \BibitemOpen
  \bibfield  {author} {\bibinfo {author} {\bibfnamefont {L.}~\bibnamefont
  {Aiello}}, \bibinfo {author} {\bibfnamefont {J.~W.}\ \bibnamefont
  {Richardson}}, \bibinfo {author} {\bibfnamefont {S.~M.}\ \bibnamefont
  {Vermeulen}}, \bibinfo {author} {\bibfnamefont {H.}~\bibnamefont {Grote}},
  \bibinfo {author} {\bibfnamefont {C.}~\bibnamefont {Hogan}}, \bibinfo
  {author} {\bibfnamefont {O.}~\bibnamefont {Kwon}}, \ and\ \bibinfo {author}
  {\bibfnamefont {C.}~\bibnamefont {Stoughton}},\ }\href@noop {} {\bibfield
  {journal} {\bibinfo  {journal} {arXiv e-prints}\ ,\ \bibinfo {pages} {arXiv}}
  (\bibinfo {year} {2021})}\BibitemShut {NoStop}%
\bibitem [{\citenamefont {Branca}\ \emph {et~al.}(2017)\citenamefont {Branca},
  \citenamefont {Bonaldi}, \citenamefont {Cerdonio}, \citenamefont {Conti},
  \citenamefont {Falferi}, \citenamefont {Marin}, \citenamefont {Mezzena},
  \citenamefont {Ortolan}, \citenamefont {Prodi}, \citenamefont {Taffarello}
  \emph {et~al.}}]{branca2017search}%
  \BibitemOpen
  \bibfield  {author} {\bibinfo {author} {\bibfnamefont {A.}~\bibnamefont
  {Branca}}, \bibinfo {author} {\bibfnamefont {M.}~\bibnamefont {Bonaldi}},
  \bibinfo {author} {\bibfnamefont {M.}~\bibnamefont {Cerdonio}}, \bibinfo
  {author} {\bibfnamefont {L.}~\bibnamefont {Conti}}, \bibinfo {author}
  {\bibfnamefont {P.}~\bibnamefont {Falferi}}, \bibinfo {author} {\bibfnamefont
  {F.}~\bibnamefont {Marin}}, \bibinfo {author} {\bibfnamefont
  {R.}~\bibnamefont {Mezzena}}, \bibinfo {author} {\bibfnamefont
  {A.}~\bibnamefont {Ortolan}}, \bibinfo {author} {\bibfnamefont {G.~A.}\
  \bibnamefont {Prodi}}, \bibinfo {author} {\bibfnamefont {L.}~\bibnamefont
  {Taffarello}},  \emph {et~al.},\ }\href@noop {} {\bibfield  {journal}
  {\bibinfo  {journal} {Physical Review Letters}\ }\textbf {\bibinfo {volume}
  {118}},\ \bibinfo {pages} {021302} (\bibinfo {year} {2017})}\BibitemShut
  {NoStop}%
\bibitem [{\citenamefont {Russell}(2006)}]{russell2006photonic}%
  \BibitemOpen
  \bibfield  {author} {\bibinfo {author} {\bibfnamefont {P.~S.~J.}\
  \bibnamefont {Russell}},\ }\href@noop {} {\bibfield  {journal} {\bibinfo
  {journal} {Journal of lightwave technology}\ }\textbf {\bibinfo {volume}
  {24}},\ \bibinfo {pages} {4729} (\bibinfo {year} {2006})}\BibitemShut
  {NoStop}%
\bibitem [{\citenamefont {Centers}\ \emph {et~al.}(2021)\citenamefont
  {Centers}, \citenamefont {Blanchard}, \citenamefont {Conrad}, \citenamefont
  {Figueroa}, \citenamefont {Garcon}, \citenamefont {Gramolin}, \citenamefont
  {Kimball}, \citenamefont {Lawson}, \citenamefont {Pelssers}, \citenamefont
  {Smiga} \emph {et~al.}}]{centers2021stochastic}%
  \BibitemOpen
  \bibfield  {author} {\bibinfo {author} {\bibfnamefont {G.~P.}\ \bibnamefont
  {Centers}}, \bibinfo {author} {\bibfnamefont {J.~W.}\ \bibnamefont
  {Blanchard}}, \bibinfo {author} {\bibfnamefont {J.}~\bibnamefont {Conrad}},
  \bibinfo {author} {\bibfnamefont {N.~L.}\ \bibnamefont {Figueroa}}, \bibinfo
  {author} {\bibfnamefont {A.}~\bibnamefont {Garcon}}, \bibinfo {author}
  {\bibfnamefont {A.~V.}\ \bibnamefont {Gramolin}}, \bibinfo {author}
  {\bibfnamefont {D.~F.~J.}\ \bibnamefont {Kimball}}, \bibinfo {author}
  {\bibfnamefont {M.}~\bibnamefont {Lawson}}, \bibinfo {author} {\bibfnamefont
  {B.}~\bibnamefont {Pelssers}}, \bibinfo {author} {\bibfnamefont {J.~A.}\
  \bibnamefont {Smiga}},  \emph {et~al.},\ }\href@noop {} {\bibfield  {journal}
  {\bibinfo  {journal} {Nature communications}\ }\textbf {\bibinfo {volume}
  {12}},\ \bibinfo {pages} {1} (\bibinfo {year} {2021})}\BibitemShut {NoStop}%
\bibitem [{\citenamefont {Derevianko}(2018)}]{derevianko2018detecting}%
  \BibitemOpen
  \bibfield  {author} {\bibinfo {author} {\bibfnamefont {A.}~\bibnamefont
  {Derevianko}},\ }\href@noop {} {\bibfield  {journal} {\bibinfo  {journal}
  {Physical Review A}\ }\textbf {\bibinfo {volume} {97}},\ \bibinfo {pages}
  {042506} (\bibinfo {year} {2018})}\BibitemShut {NoStop}%
\bibitem [{\citenamefont {Damour}\ and\ \citenamefont
  {Donoghue}(2010)}]{damour2010equivalence}%
  \BibitemOpen
  \bibfield  {author} {\bibinfo {author} {\bibfnamefont {T.}~\bibnamefont
  {Damour}}\ and\ \bibinfo {author} {\bibfnamefont {J.~F.}\ \bibnamefont
  {Donoghue}},\ }\href@noop {} {\bibfield  {journal} {\bibinfo  {journal}
  {Physical Review D}\ }\textbf {\bibinfo {volume} {82}},\ \bibinfo {pages}
  {084033} (\bibinfo {year} {2010})}\BibitemShut {NoStop}%
\bibitem [{\citenamefont {Braxmaier}\ \emph {et~al.}(2001)\citenamefont
  {Braxmaier}, \citenamefont {Pradl}, \citenamefont {M{\"u}ller}, \citenamefont
  {Peters}, \citenamefont {Mlynek}, \citenamefont {Loriette},\ and\
  \citenamefont {Schiller}}]{braxmaier2001proposed}%
  \BibitemOpen
  \bibfield  {author} {\bibinfo {author} {\bibfnamefont {C.}~\bibnamefont
  {Braxmaier}}, \bibinfo {author} {\bibfnamefont {O.}~\bibnamefont {Pradl}},
  \bibinfo {author} {\bibfnamefont {H.}~\bibnamefont {M{\"u}ller}}, \bibinfo
  {author} {\bibfnamefont {A.}~\bibnamefont {Peters}}, \bibinfo {author}
  {\bibfnamefont {J.}~\bibnamefont {Mlynek}}, \bibinfo {author} {\bibfnamefont
  {V.}~\bibnamefont {Loriette}}, \ and\ \bibinfo {author} {\bibfnamefont
  {S.}~\bibnamefont {Schiller}},\ }\href@noop {} {\bibfield  {journal}
  {\bibinfo  {journal} {Physical Review D}\ }\textbf {\bibinfo {volume} {64}},\
  \bibinfo {pages} {042001} (\bibinfo {year} {2001})}\BibitemShut {NoStop}%
\bibitem [{\citenamefont {Stadnik}\ and\ \citenamefont
  {Flambaum}(2015)}]{stadnik2015searching}%
  \BibitemOpen
  \bibfield  {author} {\bibinfo {author} {\bibfnamefont {Y.}~\bibnamefont
  {Stadnik}}\ and\ \bibinfo {author} {\bibfnamefont {V.}~\bibnamefont
  {Flambaum}},\ }\href@noop {} {\bibfield  {journal} {\bibinfo  {journal}
  {Physical review letters}\ }\textbf {\bibinfo {volume} {114}},\ \bibinfo
  {pages} {161301} (\bibinfo {year} {2015})}\BibitemShut {NoStop}%
\bibitem [{\citenamefont {Arvanitaki}\ \emph {et~al.}(2016)\citenamefont
  {Arvanitaki}, \citenamefont {Dimopoulos},\ and\ \citenamefont
  {Van~Tilburg}}]{arvanitaki2016sound}%
  \BibitemOpen
  \bibfield  {author} {\bibinfo {author} {\bibfnamefont {A.}~\bibnamefont
  {Arvanitaki}}, \bibinfo {author} {\bibfnamefont {S.}~\bibnamefont
  {Dimopoulos}}, \ and\ \bibinfo {author} {\bibfnamefont {K.}~\bibnamefont
  {Van~Tilburg}},\ }\href@noop {} {\bibfield  {journal} {\bibinfo  {journal}
  {Physical review letters}\ }\textbf {\bibinfo {volume} {116}},\ \bibinfo
  {pages} {031102} (\bibinfo {year} {2016})}\BibitemShut {NoStop}%
\bibitem [{\citenamefont {Grote}\ and\ \citenamefont
  {Stadnik}(2019)}]{grote2019novel}%
  \BibitemOpen
  \bibfield  {author} {\bibinfo {author} {\bibfnamefont {H.}~\bibnamefont
  {Grote}}\ and\ \bibinfo {author} {\bibfnamefont {Y.}~\bibnamefont
  {Stadnik}},\ }\href@noop {} {\bibfield  {journal} {\bibinfo  {journal}
  {Physical Review Research}\ }\textbf {\bibinfo {volume} {1}},\ \bibinfo
  {pages} {033187} (\bibinfo {year} {2019})}\BibitemShut {NoStop}%
\bibitem [{\citenamefont {Shi}\ \emph {et~al.}(2021)\citenamefont {Shi},
  \citenamefont {Sakr}, \citenamefont {Hayes}, \citenamefont {Wei},
  \citenamefont {Fokoua}, \citenamefont {Ding}, \citenamefont {Feng},
  \citenamefont {Marra}, \citenamefont {Poletti}, \citenamefont {Richardson}
  \emph {et~al.}}]{shi2021thinly}%
  \BibitemOpen
  \bibfield  {author} {\bibinfo {author} {\bibfnamefont {B.}~\bibnamefont
  {Shi}}, \bibinfo {author} {\bibfnamefont {H.}~\bibnamefont {Sakr}}, \bibinfo
  {author} {\bibfnamefont {J.}~\bibnamefont {Hayes}}, \bibinfo {author}
  {\bibfnamefont {X.}~\bibnamefont {Wei}}, \bibinfo {author} {\bibfnamefont
  {E.~N.}\ \bibnamefont {Fokoua}}, \bibinfo {author} {\bibfnamefont
  {M.}~\bibnamefont {Ding}}, \bibinfo {author} {\bibfnamefont {Z.}~\bibnamefont
  {Feng}}, \bibinfo {author} {\bibfnamefont {G.}~\bibnamefont {Marra}},
  \bibinfo {author} {\bibfnamefont {F.}~\bibnamefont {Poletti}}, \bibinfo
  {author} {\bibfnamefont {D.~J.}\ \bibnamefont {Richardson}},  \emph
  {et~al.},\ }\href@noop {} {\bibfield  {journal} {\bibinfo  {journal} {Optics
  Letters}\ }\textbf {\bibinfo {volume} {46}},\ \bibinfo {pages} {5177}
  (\bibinfo {year} {2021})}\BibitemShut {NoStop}%
\bibitem [{\citenamefont {Bradley}\ \emph {et~al.}(2019)\citenamefont
  {Bradley}, \citenamefont {Jasion}, \citenamefont {Hayes}, \citenamefont
  {Chen}, \citenamefont {Hooper}, \citenamefont {Sakr}, \citenamefont {Alonso},
  \citenamefont {Taranta}, \citenamefont {Saljoghei}, \citenamefont {Mulvad}
  \emph {et~al.}}]{bradley2019antiresonant}%
  \BibitemOpen
  \bibfield  {author} {\bibinfo {author} {\bibfnamefont {T.~D.}\ \bibnamefont
  {Bradley}}, \bibinfo {author} {\bibfnamefont {G.~T.}\ \bibnamefont {Jasion}},
  \bibinfo {author} {\bibfnamefont {J.~R.}\ \bibnamefont {Hayes}}, \bibinfo
  {author} {\bibfnamefont {Y.}~\bibnamefont {Chen}}, \bibinfo {author}
  {\bibfnamefont {L.}~\bibnamefont {Hooper}}, \bibinfo {author} {\bibfnamefont
  {H.}~\bibnamefont {Sakr}}, \bibinfo {author} {\bibfnamefont {M.}~\bibnamefont
  {Alonso}}, \bibinfo {author} {\bibfnamefont {A.}~\bibnamefont {Taranta}},
  \bibinfo {author} {\bibfnamefont {A.}~\bibnamefont {Saljoghei}}, \bibinfo
  {author} {\bibfnamefont {H.~C.}\ \bibnamefont {Mulvad}},  \emph {et~al.},\
  }in\ \href@noop {} {\emph {\bibinfo {booktitle} {45th European Conference on
  Optical Communication (ECOC 2019)}}}\ (\bibinfo {organization} {IET},\
  \bibinfo {year} {2019})\ pp.\ \bibinfo {pages} {1--4}\BibitemShut {NoStop}%
\bibitem [{\citenamefont {Jasion}\ \emph {et~al.}(2021)\citenamefont {Jasion},
  \citenamefont {Bradley}, \citenamefont {Harrington}, \citenamefont {Sakr},
  \citenamefont {Chen}, \citenamefont {Fokoua}, \citenamefont {Davidson},
  \citenamefont {Taranta}, \citenamefont {Hayes}, \citenamefont {Richardson}
  \emph {et~al.}}]{jasion2021recent}%
  \BibitemOpen
  \bibfield  {author} {\bibinfo {author} {\bibfnamefont {G.~T.}\ \bibnamefont
  {Jasion}}, \bibinfo {author} {\bibfnamefont {T.~D.}\ \bibnamefont {Bradley}},
  \bibinfo {author} {\bibfnamefont {K.}~\bibnamefont {Harrington}}, \bibinfo
  {author} {\bibfnamefont {H.}~\bibnamefont {Sakr}}, \bibinfo {author}
  {\bibfnamefont {Y.}~\bibnamefont {Chen}}, \bibinfo {author} {\bibfnamefont
  {E.~N.}\ \bibnamefont {Fokoua}}, \bibinfo {author} {\bibfnamefont {I.~A.}\
  \bibnamefont {Davidson}}, \bibinfo {author} {\bibfnamefont {A.}~\bibnamefont
  {Taranta}}, \bibinfo {author} {\bibfnamefont {J.~R.}\ \bibnamefont {Hayes}},
  \bibinfo {author} {\bibfnamefont {D.~J.}\ \bibnamefont {Richardson}},  \emph
  {et~al.},\ }in\ \href@noop {} {\emph {\bibinfo {booktitle} {Optical Fiber
  Communication Conference}}}\ (\bibinfo {organization} {Optical Society of
  America},\ \bibinfo {year} {2021})\ pp.\ \bibinfo {pages}
  {M5E--2}\BibitemShut {NoStop}%
\bibitem [{\citenamefont {Pa{\v{s}}teka}\ \emph {et~al.}(2019)\citenamefont
  {Pa{\v{s}}teka}, \citenamefont {Hao}, \citenamefont {Borschevsky},
  \citenamefont {Flambaum},\ and\ \citenamefont
  {Schwerdtfeger}}]{pavsteka2019material}%
  \BibitemOpen
  \bibfield  {author} {\bibinfo {author} {\bibfnamefont {L.~F.}\ \bibnamefont
  {Pa{\v{s}}teka}}, \bibinfo {author} {\bibfnamefont {Y.}~\bibnamefont {Hao}},
  \bibinfo {author} {\bibfnamefont {A.}~\bibnamefont {Borschevsky}}, \bibinfo
  {author} {\bibfnamefont {V.~V.}\ \bibnamefont {Flambaum}}, \ and\ \bibinfo
  {author} {\bibfnamefont {P.}~\bibnamefont {Schwerdtfeger}},\ }\href@noop {}
  {\bibfield  {journal} {\bibinfo  {journal} {Physical Review Letters}\
  }\textbf {\bibinfo {volume} {122}},\ \bibinfo {pages} {160801} (\bibinfo
  {year} {2019})}\BibitemShut {NoStop}%
\bibitem [{\citenamefont {Duan}(2012)}]{duan2012general}%
  \BibitemOpen
  \bibfield  {author} {\bibinfo {author} {\bibfnamefont {L.}~\bibnamefont
  {Duan}},\ }\href@noop {} {\bibfield  {journal} {\bibinfo  {journal} {Physical
  Review A}\ }\textbf {\bibinfo {volume} {86}},\ \bibinfo {pages} {023817}
  (\bibinfo {year} {2012})}\BibitemShut {NoStop}%
\bibitem [{\citenamefont {Duan}(2010)}]{duan2010intrinsic}%
  \BibitemOpen
  \bibfield  {author} {\bibinfo {author} {\bibfnamefont {L.}~\bibnamefont
  {Duan}},\ }\href@noop {} {\bibfield  {journal} {\bibinfo  {journal}
  {Electronics letters}\ }\textbf {\bibinfo {volume} {46}},\ \bibinfo {pages}
  {1515} (\bibinfo {year} {2010})}\BibitemShut {NoStop}%
\bibitem [{\citenamefont {Bartolo}\ \emph {et~al.}(2012)\citenamefont
  {Bartolo}, \citenamefont {Tveten},\ and\ \citenamefont
  {Dandridge}}]{bartolo2012thermal}%
  \BibitemOpen
  \bibfield  {author} {\bibinfo {author} {\bibfnamefont {R.~E.}\ \bibnamefont
  {Bartolo}}, \bibinfo {author} {\bibfnamefont {A.~B.}\ \bibnamefont {Tveten}},
  \ and\ \bibinfo {author} {\bibfnamefont {A.}~\bibnamefont {Dandridge}},\
  }\href@noop {} {\bibfield  {journal} {\bibinfo  {journal} {IEEE Journal of
  Quantum Electronics}\ }\textbf {\bibinfo {volume} {48}},\ \bibinfo {pages}
  {720} (\bibinfo {year} {2012})}\BibitemShut {NoStop}%
\bibitem [{\citenamefont {Dong}\ \emph {et~al.}(2016)\citenamefont {Dong},
  \citenamefont {Huang}, \citenamefont {Li},\ and\ \citenamefont
  {Liu}}]{dong2016observation}%
  \BibitemOpen
  \bibfield  {author} {\bibinfo {author} {\bibfnamefont {J.}~\bibnamefont
  {Dong}}, \bibinfo {author} {\bibfnamefont {J.}~\bibnamefont {Huang}},
  \bibinfo {author} {\bibfnamefont {T.}~\bibnamefont {Li}}, \ and\ \bibinfo
  {author} {\bibfnamefont {L.}~\bibnamefont {Liu}},\ }\href@noop {} {\bibfield
  {journal} {\bibinfo  {journal} {Applied Physics Letters}\ }\textbf {\bibinfo
  {volume} {108}},\ \bibinfo {pages} {021108} (\bibinfo {year}
  {2016})}\BibitemShut {NoStop}%
\bibitem [{\citenamefont {Huang}\ \emph
  {et~al.}(2019{\natexlab{a}})\citenamefont {Huang}, \citenamefont {Wang},
  \citenamefont {Duan}, \citenamefont {Huang}, \citenamefont {Ye},
  \citenamefont {Liu},\ and\ \citenamefont {Li}}]{huang2019all}%
  \BibitemOpen
  \bibfield  {author} {\bibinfo {author} {\bibfnamefont {J.}~\bibnamefont
  {Huang}}, \bibinfo {author} {\bibfnamefont {L.}~\bibnamefont {Wang}},
  \bibinfo {author} {\bibfnamefont {Y.}~\bibnamefont {Duan}}, \bibinfo {author}
  {\bibfnamefont {Y.}~\bibnamefont {Huang}}, \bibinfo {author} {\bibfnamefont
  {M.}~\bibnamefont {Ye}}, \bibinfo {author} {\bibfnamefont {L.}~\bibnamefont
  {Liu}}, \ and\ \bibinfo {author} {\bibfnamefont {T.}~\bibnamefont {Li}},\
  }\href@noop {} {\bibfield  {journal} {\bibinfo  {journal} {Chinese Optics
  Letters}\ }\textbf {\bibinfo {volume} {17}},\ \bibinfo {pages} {071407}
  (\bibinfo {year} {2019}{\natexlab{a}})}\BibitemShut {NoStop}%
\bibitem [{\citenamefont {Li}\ \emph {et~al.}(2011)\citenamefont {Li},
  \citenamefont {Argence}, \citenamefont {Haboucha}, \citenamefont {Jiang},
  \citenamefont {Dornaux}, \citenamefont {Kon{\'e}}, \citenamefont {Clairon},
  \citenamefont {Lemonde}, \citenamefont {Santarelli}, \citenamefont {Nelson}
  \emph {et~al.}}]{li2011low}%
  \BibitemOpen
  \bibfield  {author} {\bibinfo {author} {\bibfnamefont {T.}~\bibnamefont
  {Li}}, \bibinfo {author} {\bibfnamefont {B.}~\bibnamefont {Argence}},
  \bibinfo {author} {\bibfnamefont {A.}~\bibnamefont {Haboucha}}, \bibinfo
  {author} {\bibfnamefont {H.}~\bibnamefont {Jiang}}, \bibinfo {author}
  {\bibfnamefont {J.}~\bibnamefont {Dornaux}}, \bibinfo {author} {\bibfnamefont
  {D.}~\bibnamefont {Kon{\'e}}}, \bibinfo {author} {\bibfnamefont
  {A.}~\bibnamefont {Clairon}}, \bibinfo {author} {\bibfnamefont
  {P.}~\bibnamefont {Lemonde}}, \bibinfo {author} {\bibfnamefont
  {G.}~\bibnamefont {Santarelli}}, \bibinfo {author} {\bibfnamefont
  {C.}~\bibnamefont {Nelson}},  \emph {et~al.},\ }in\ \href@noop {} {\emph
  {\bibinfo {booktitle} {2011 Joint Conference of the IEEE International
  Frequency Control and the European Frequency and Time Forum (FCS)
  Proceedings}}}\ (\bibinfo {organization} {IEEE},\ \bibinfo {year} {2011})\
  pp.\ \bibinfo {pages} {1--3}\BibitemShut {NoStop}%
\bibitem [{\citenamefont {Peterson}(1993)}]{peterson1993observations}%
  \BibitemOpen
  \bibfield  {author} {\bibinfo {author} {\bibfnamefont {J.~R.}\ \bibnamefont
  {Peterson}},\ }\href@noop {} {\emph {\bibinfo {title} {Observations and
  modeling of seismic background noise}}},\ \bibinfo {type} {Tech. Rep.}\
  (\bibinfo  {institution} {US Geological Survey},\ \bibinfo {year}
  {1993})\BibitemShut {NoStop}%
\bibitem [{\citenamefont {Berg{\'e}}\ \emph {et~al.}(2018)\citenamefont
  {Berg{\'e}}, \citenamefont {Brax}, \citenamefont {M{\'e}tris}, \citenamefont
  {Pernot-Borr{\`a}s}, \citenamefont {Touboul},\ and\ \citenamefont
  {Uzan}}]{berge2018microscope}%
  \BibitemOpen
  \bibfield  {author} {\bibinfo {author} {\bibfnamefont {J.}~\bibnamefont
  {Berg{\'e}}}, \bibinfo {author} {\bibfnamefont {P.}~\bibnamefont {Brax}},
  \bibinfo {author} {\bibfnamefont {G.}~\bibnamefont {M{\'e}tris}}, \bibinfo
  {author} {\bibfnamefont {M.}~\bibnamefont {Pernot-Borr{\`a}s}}, \bibinfo
  {author} {\bibfnamefont {P.}~\bibnamefont {Touboul}}, \ and\ \bibinfo
  {author} {\bibfnamefont {J.-P.}\ \bibnamefont {Uzan}},\ }\href@noop {}
  {\bibfield  {journal} {\bibinfo  {journal} {Physical review letters}\
  }\textbf {\bibinfo {volume} {120}},\ \bibinfo {pages} {141101} (\bibinfo
  {year} {2018})}\BibitemShut {NoStop}%
\bibitem [{\citenamefont {Jasion}\ \emph {et~al.}(2022)\citenamefont {Jasion},
  \citenamefont {Sakr}, \citenamefont {Hayes}, \citenamefont {Sandoghchi},
  \citenamefont {Hooper}, \citenamefont {Fokoua}, \citenamefont {Saljoghei},
  \citenamefont {Mulvad}, \citenamefont {Alonso}, \citenamefont {Taranta} \emph
  {et~al.}}]{jasion20220}%
  \BibitemOpen
  \bibfield  {author} {\bibinfo {author} {\bibfnamefont {G.~T.}\ \bibnamefont
  {Jasion}}, \bibinfo {author} {\bibfnamefont {H.}~\bibnamefont {Sakr}},
  \bibinfo {author} {\bibfnamefont {J.~R.}\ \bibnamefont {Hayes}}, \bibinfo
  {author} {\bibfnamefont {S.~R.}\ \bibnamefont {Sandoghchi}}, \bibinfo
  {author} {\bibfnamefont {L.}~\bibnamefont {Hooper}}, \bibinfo {author}
  {\bibfnamefont {E.~N.}\ \bibnamefont {Fokoua}}, \bibinfo {author}
  {\bibfnamefont {A.}~\bibnamefont {Saljoghei}}, \bibinfo {author}
  {\bibfnamefont {H.~C.}\ \bibnamefont {Mulvad}}, \bibinfo {author}
  {\bibfnamefont {M.}~\bibnamefont {Alonso}}, \bibinfo {author} {\bibfnamefont
  {A.}~\bibnamefont {Taranta}},  \emph {et~al.},\ }in\ \href@noop {} {\emph
  {\bibinfo {booktitle} {2022 Optical Fiber Communications Conference and
  Exhibition (OFC)}}}\ (\bibinfo {organization} {IEEE},\ \bibinfo {year}
  {2022})\ pp.\ \bibinfo {pages} {1--3}\BibitemShut {NoStop}%
\bibitem [{smf(2014)}]{smf28}%
  \BibitemOpen
  \href
  {https://www.corning.com/media/worldwide/coc/documents/Fiber/SMF-28%20Ultra.pdf}
  {\emph {\bibinfo {title} {Corning SMF-28 Ultra Optical Fiber Product
  Information}}},\ \bibinfo {organization} {Corning} (\bibinfo {year}
  {2014})\BibitemShut {NoStop}%
\bibitem [{\citenamefont {Hilweg}\ \emph {et~al.}(2022)\citenamefont {Hilweg},
  \citenamefont {Shadmany}, \citenamefont {Walther}, \citenamefont
  {Mavalvala},\ and\ \citenamefont {Sudhir}}]{hilweg2022limits}%
  \BibitemOpen
  \bibfield  {author} {\bibinfo {author} {\bibfnamefont {C.}~\bibnamefont
  {Hilweg}}, \bibinfo {author} {\bibfnamefont {D.}~\bibnamefont {Shadmany}},
  \bibinfo {author} {\bibfnamefont {P.}~\bibnamefont {Walther}}, \bibinfo
  {author} {\bibfnamefont {N.}~\bibnamefont {Mavalvala}}, \ and\ \bibinfo
  {author} {\bibfnamefont {V.}~\bibnamefont {Sudhir}},\ }\href@noop {}
  {\bibfield  {journal} {\bibinfo  {journal} {Optica}\ }\textbf {\bibinfo
  {volume} {9}},\ \bibinfo {pages} {1238} (\bibinfo {year} {2022})}\BibitemShut
  {NoStop}%
\bibitem [{\citenamefont {Budker}\ \emph {et~al.}(2014)\citenamefont {Budker},
  \citenamefont {Graham}, \citenamefont {Ledbetter}, \citenamefont
  {Rajendran},\ and\ \citenamefont {Sushkov}}]{budker2014proposal}%
  \BibitemOpen
  \bibfield  {author} {\bibinfo {author} {\bibfnamefont {D.}~\bibnamefont
  {Budker}}, \bibinfo {author} {\bibfnamefont {P.~W.}\ \bibnamefont {Graham}},
  \bibinfo {author} {\bibfnamefont {M.}~\bibnamefont {Ledbetter}}, \bibinfo
  {author} {\bibfnamefont {S.}~\bibnamefont {Rajendran}}, \ and\ \bibinfo
  {author} {\bibfnamefont {A.~O.}\ \bibnamefont {Sushkov}},\ }\href@noop {}
  {\bibfield  {journal} {\bibinfo  {journal} {Physical Review X}\ }\textbf
  {\bibinfo {volume} {4}},\ \bibinfo {pages} {021030} (\bibinfo {year}
  {2014})}\BibitemShut {NoStop}%
\bibitem [{\citenamefont {Banerjee}\ \emph
  {et~al.}(2020{\natexlab{a}})\citenamefont {Banerjee}, \citenamefont {Budker},
  \citenamefont {Eby}, \citenamefont {Kim},\ and\ \citenamefont
  {Perez}}]{banerjee2020relaxion}%
  \BibitemOpen
  \bibfield  {author} {\bibinfo {author} {\bibfnamefont {A.}~\bibnamefont
  {Banerjee}}, \bibinfo {author} {\bibfnamefont {D.}~\bibnamefont {Budker}},
  \bibinfo {author} {\bibfnamefont {J.}~\bibnamefont {Eby}}, \bibinfo {author}
  {\bibfnamefont {H.}~\bibnamefont {Kim}}, \ and\ \bibinfo {author}
  {\bibfnamefont {G.}~\bibnamefont {Perez}},\ }\href@noop {} {\bibfield
  {journal} {\bibinfo  {journal} {Communications Physics}\ }\textbf {\bibinfo
  {volume} {3}},\ \bibinfo {pages} {1} (\bibinfo {year}
  {2020}{\natexlab{a}})}\BibitemShut {NoStop}%
\bibitem [{\citenamefont {Oswald}\ \emph {et~al.}(2021)\citenamefont {Oswald},
  \citenamefont {Nevsky}, \citenamefont {Vogt}, \citenamefont {Schiller},
  \citenamefont {Figueroa}, \citenamefont {Zhang}, \citenamefont {Tretiak},
  \citenamefont {Antypas}, \citenamefont {Budker}, \citenamefont {Banerjee}
  \emph {et~al.}}]{oswald2021search}%
  \BibitemOpen
  \bibfield  {author} {\bibinfo {author} {\bibfnamefont {R.}~\bibnamefont
  {Oswald}}, \bibinfo {author} {\bibfnamefont {A.}~\bibnamefont {Nevsky}},
  \bibinfo {author} {\bibfnamefont {V.}~\bibnamefont {Vogt}}, \bibinfo {author}
  {\bibfnamefont {S.}~\bibnamefont {Schiller}}, \bibinfo {author}
  {\bibfnamefont {N.}~\bibnamefont {Figueroa}}, \bibinfo {author}
  {\bibfnamefont {K.}~\bibnamefont {Zhang}}, \bibinfo {author} {\bibfnamefont
  {O.}~\bibnamefont {Tretiak}}, \bibinfo {author} {\bibfnamefont
  {D.}~\bibnamefont {Antypas}}, \bibinfo {author} {\bibfnamefont
  {D.}~\bibnamefont {Budker}}, \bibinfo {author} {\bibfnamefont
  {A.}~\bibnamefont {Banerjee}},  \emph {et~al.},\ }\href@noop {} {\bibfield
  {journal} {\bibinfo  {journal} {arXiv preprint arXiv:2111.06883}\ } (\bibinfo
  {year} {2021})}\BibitemShut {NoStop}%
\bibitem [{\citenamefont {Antypas}\ \emph {et~al.}(2019)\citenamefont
  {Antypas}, \citenamefont {Tretiak}, \citenamefont {Garcon}, \citenamefont
  {Ozeri}, \citenamefont {Perez},\ and\ \citenamefont
  {Budker}}]{antypas2019scalar}%
  \BibitemOpen
  \bibfield  {author} {\bibinfo {author} {\bibfnamefont {D.}~\bibnamefont
  {Antypas}}, \bibinfo {author} {\bibfnamefont {O.}~\bibnamefont {Tretiak}},
  \bibinfo {author} {\bibfnamefont {A.}~\bibnamefont {Garcon}}, \bibinfo
  {author} {\bibfnamefont {R.}~\bibnamefont {Ozeri}}, \bibinfo {author}
  {\bibfnamefont {G.}~\bibnamefont {Perez}}, \ and\ \bibinfo {author}
  {\bibfnamefont {D.}~\bibnamefont {Budker}},\ }\href@noop {} {\bibfield
  {journal} {\bibinfo  {journal} {Physical Review Letters}\ }\textbf {\bibinfo
  {volume} {123}},\ \bibinfo {pages} {141102} (\bibinfo {year}
  {2019})}\BibitemShut {NoStop}%
\bibitem [{\citenamefont {Aharony}\ \emph {et~al.}(2021)\citenamefont
  {Aharony}, \citenamefont {Akerman}, \citenamefont {Ozeri}, \citenamefont
  {Perez}, \citenamefont {Savoray},\ and\ \citenamefont
  {Shaniv}}]{aharony2021constraining}%
  \BibitemOpen
  \bibfield  {author} {\bibinfo {author} {\bibfnamefont {S.}~\bibnamefont
  {Aharony}}, \bibinfo {author} {\bibfnamefont {N.}~\bibnamefont {Akerman}},
  \bibinfo {author} {\bibfnamefont {R.}~\bibnamefont {Ozeri}}, \bibinfo
  {author} {\bibfnamefont {G.}~\bibnamefont {Perez}}, \bibinfo {author}
  {\bibfnamefont {I.}~\bibnamefont {Savoray}}, \ and\ \bibinfo {author}
  {\bibfnamefont {R.}~\bibnamefont {Shaniv}},\ }\href@noop {} {\bibfield
  {journal} {\bibinfo  {journal} {Physical Review D}\ }\textbf {\bibinfo
  {volume} {103}},\ \bibinfo {pages} {075017} (\bibinfo {year}
  {2021})}\BibitemShut {NoStop}%
\bibitem [{\citenamefont {Banerjee}\ \emph
  {et~al.}(2020{\natexlab{b}})\citenamefont {Banerjee}, \citenamefont {Budker},
  \citenamefont {Eby}, \citenamefont {Flambaum}, \citenamefont {Kim},
  \citenamefont {Matsedonskyi},\ and\ \citenamefont
  {Perez}}]{banerjee2020searching}%
  \BibitemOpen
  \bibfield  {author} {\bibinfo {author} {\bibfnamefont {A.}~\bibnamefont
  {Banerjee}}, \bibinfo {author} {\bibfnamefont {D.}~\bibnamefont {Budker}},
  \bibinfo {author} {\bibfnamefont {J.}~\bibnamefont {Eby}}, \bibinfo {author}
  {\bibfnamefont {V.~V.}\ \bibnamefont {Flambaum}}, \bibinfo {author}
  {\bibfnamefont {H.}~\bibnamefont {Kim}}, \bibinfo {author} {\bibfnamefont
  {O.}~\bibnamefont {Matsedonskyi}}, \ and\ \bibinfo {author} {\bibfnamefont
  {G.}~\bibnamefont {Perez}},\ }\href@noop {} {\bibfield  {journal} {\bibinfo
  {journal} {Journal of High Energy Physics}\ }\textbf {\bibinfo {volume}
  {2020}},\ \bibinfo {pages} {1} (\bibinfo {year}
  {2020}{\natexlab{b}})}\BibitemShut {NoStop}%
\bibitem [{\citenamefont {Schlamminger}\ \emph {et~al.}(2008)\citenamefont
  {Schlamminger}, \citenamefont {Choi}, \citenamefont {Wagner}, \citenamefont
  {Gundlach},\ and\ \citenamefont {Adelberger}}]{schlamminger2008test}%
  \BibitemOpen
  \bibfield  {author} {\bibinfo {author} {\bibfnamefont {S.}~\bibnamefont
  {Schlamminger}}, \bibinfo {author} {\bibfnamefont {K.-Y.}\ \bibnamefont
  {Choi}}, \bibinfo {author} {\bibfnamefont {T.~A.}\ \bibnamefont {Wagner}},
  \bibinfo {author} {\bibfnamefont {J.~H.}\ \bibnamefont {Gundlach}}, \ and\
  \bibinfo {author} {\bibfnamefont {E.~G.}\ \bibnamefont {Adelberger}},\
  }\href@noop {} {\bibfield  {journal} {\bibinfo  {journal} {Physical Review
  Letters}\ }\textbf {\bibinfo {volume} {100}},\ \bibinfo {pages} {041101}
  (\bibinfo {year} {2008})}\BibitemShut {NoStop}%
\bibitem [{\citenamefont {Carney}\ \emph {et~al.}(2021)\citenamefont {Carney},
  \citenamefont {Hook}, \citenamefont {Liu}, \citenamefont {Taylor},\ and\
  \citenamefont {Zhao}}]{carney2021ultralight}%
  \BibitemOpen
  \bibfield  {author} {\bibinfo {author} {\bibfnamefont {D.}~\bibnamefont
  {Carney}}, \bibinfo {author} {\bibfnamefont {A.}~\bibnamefont {Hook}},
  \bibinfo {author} {\bibfnamefont {Z.}~\bibnamefont {Liu}}, \bibinfo {author}
  {\bibfnamefont {J.~M.}\ \bibnamefont {Taylor}}, \ and\ \bibinfo {author}
  {\bibfnamefont {Y.}~\bibnamefont {Zhao}},\ }\href@noop {} {\bibfield
  {journal} {\bibinfo  {journal} {New Journal of Physics}\ }\textbf {\bibinfo
  {volume} {23}},\ \bibinfo {pages} {023041} (\bibinfo {year}
  {2021})}\BibitemShut {NoStop}%
\bibitem [{\citenamefont {Delva}\ \emph {et~al.}(2017)\citenamefont {Delva},
  \citenamefont {Lodewyck}, \citenamefont {Bilicki}, \citenamefont {Bookjans},
  \citenamefont {Vallet}, \citenamefont {Le~Targat}, \citenamefont {Pottie},
  \citenamefont {Guerlin}, \citenamefont {Meynadier}, \citenamefont
  {Le~Poncin-Lafitte} \emph {et~al.}}]{delva2017test}%
  \BibitemOpen
  \bibfield  {author} {\bibinfo {author} {\bibfnamefont {P.}~\bibnamefont
  {Delva}}, \bibinfo {author} {\bibfnamefont {J.}~\bibnamefont {Lodewyck}},
  \bibinfo {author} {\bibfnamefont {S.}~\bibnamefont {Bilicki}}, \bibinfo
  {author} {\bibfnamefont {E.}~\bibnamefont {Bookjans}}, \bibinfo {author}
  {\bibfnamefont {G.}~\bibnamefont {Vallet}}, \bibinfo {author} {\bibfnamefont
  {R.}~\bibnamefont {Le~Targat}}, \bibinfo {author} {\bibfnamefont {P.-E.}\
  \bibnamefont {Pottie}}, \bibinfo {author} {\bibfnamefont {C.}~\bibnamefont
  {Guerlin}}, \bibinfo {author} {\bibfnamefont {F.}~\bibnamefont {Meynadier}},
  \bibinfo {author} {\bibfnamefont {C.}~\bibnamefont {Le~Poncin-Lafitte}},
  \emph {et~al.},\ }\href@noop {} {\bibfield  {journal} {\bibinfo  {journal}
  {Physical review letters}\ }\textbf {\bibinfo {volume} {118}},\ \bibinfo
  {pages} {221102} (\bibinfo {year} {2017})}\BibitemShut {NoStop}%
\bibitem [{\citenamefont {Lisdat}\ \emph {et~al.}(2016)\citenamefont {Lisdat},
  \citenamefont {Grosche}, \citenamefont {Quintin}, \citenamefont {Shi},
  \citenamefont {Raupach}, \citenamefont {Grebing}, \citenamefont {Nicolodi},
  \citenamefont {Stefani}, \citenamefont {Al-Masoudi}, \citenamefont
  {D{\"o}rscher} \emph {et~al.}}]{lisdat2016clock}%
  \BibitemOpen
  \bibfield  {author} {\bibinfo {author} {\bibfnamefont {C.}~\bibnamefont
  {Lisdat}}, \bibinfo {author} {\bibfnamefont {G.}~\bibnamefont {Grosche}},
  \bibinfo {author} {\bibfnamefont {N.}~\bibnamefont {Quintin}}, \bibinfo
  {author} {\bibfnamefont {C.}~\bibnamefont {Shi}}, \bibinfo {author}
  {\bibfnamefont {S.}~\bibnamefont {Raupach}}, \bibinfo {author} {\bibfnamefont
  {C.}~\bibnamefont {Grebing}}, \bibinfo {author} {\bibfnamefont
  {D.}~\bibnamefont {Nicolodi}}, \bibinfo {author} {\bibfnamefont
  {F.}~\bibnamefont {Stefani}}, \bibinfo {author} {\bibfnamefont
  {A.}~\bibnamefont {Al-Masoudi}}, \bibinfo {author} {\bibfnamefont
  {S.}~\bibnamefont {D{\"o}rscher}},  \emph {et~al.},\ }\href@noop {}
  {\bibfield  {journal} {\bibinfo  {journal} {Nature communications}\ }\textbf
  {\bibinfo {volume} {7}},\ \bibinfo {pages} {12443} (\bibinfo {year}
  {2016})}\BibitemShut {NoStop}%
\bibitem [{\citenamefont {Yavuz}\ and\ \citenamefont
  {Inbar}(2022)}]{yavuz2022generation}%
  \BibitemOpen
  \bibfield  {author} {\bibinfo {author} {\bibfnamefont {D.}~\bibnamefont
  {Yavuz}}\ and\ \bibinfo {author} {\bibfnamefont {S.}~\bibnamefont {Inbar}},\
  }\href@noop {} {\bibfield  {journal} {\bibinfo  {journal} {Physical Review
  D}\ }\textbf {\bibinfo {volume} {105}},\ \bibinfo {pages} {074012} (\bibinfo
  {year} {2022})}\BibitemShut {NoStop}%
\bibitem [{\citenamefont {Kozlov}\ and\ \citenamefont
  {Budker}(2019)}]{kozlov2019comment}%
  \BibitemOpen
  \bibfield  {author} {\bibinfo {author} {\bibfnamefont {M.~G.}\ \bibnamefont
  {Kozlov}}\ and\ \bibinfo {author} {\bibfnamefont {D.}~\bibnamefont
  {Budker}},\ }\href@noop {} {\bibfield  {journal} {\bibinfo  {journal}
  {Annalen der Physik}\ }\textbf {\bibinfo {volume} {531}},\ \bibinfo {pages}
  {1800254} (\bibinfo {year} {2019})}\BibitemShut {NoStop}%
\bibitem [{\citenamefont {Malitson}(1965)}]{malitson1965interspecimen}%
  \BibitemOpen
  \bibfield  {author} {\bibinfo {author} {\bibfnamefont {I.~H.}\ \bibnamefont
  {Malitson}},\ }\href@noop {} {\bibfield  {journal} {\bibinfo  {journal}
  {Josa}\ }\textbf {\bibinfo {volume} {55}},\ \bibinfo {pages} {1205} (\bibinfo
  {year} {1965})}\BibitemShut {NoStop}%
\bibitem [{\citenamefont {Wu}\ \emph {et~al.}(2016)\citenamefont {Wu},
  \citenamefont {Jiang}, \citenamefont {Ma}, \citenamefont {Qi}, \citenamefont
  {Yu}, \citenamefont {Bi},\ and\ \citenamefont {Ma}}]{wu20160}%
  \BibitemOpen
  \bibfield  {author} {\bibinfo {author} {\bibfnamefont {L.}~\bibnamefont
  {Wu}}, \bibinfo {author} {\bibfnamefont {Y.}~\bibnamefont {Jiang}}, \bibinfo
  {author} {\bibfnamefont {C.}~\bibnamefont {Ma}}, \bibinfo {author}
  {\bibfnamefont {W.}~\bibnamefont {Qi}}, \bibinfo {author} {\bibfnamefont
  {H.}~\bibnamefont {Yu}}, \bibinfo {author} {\bibfnamefont {Z.}~\bibnamefont
  {Bi}}, \ and\ \bibinfo {author} {\bibfnamefont {L.}~\bibnamefont {Ma}},\
  }\href@noop {} {\bibfield  {journal} {\bibinfo  {journal} {Scientific
  Reports}\ }\textbf {\bibinfo {volume} {6}},\ \bibinfo {pages} {1} (\bibinfo
  {year} {2016})}\BibitemShut {NoStop}%
\bibitem [{\citenamefont {Dong}\ \emph {et~al.}(2015)\citenamefont {Dong},
  \citenamefont {Hu}, \citenamefont {Huang}, \citenamefont {Ye}, \citenamefont
  {Qu}, \citenamefont {Li},\ and\ \citenamefont {Liu}}]{dong2015subhertz}%
  \BibitemOpen
  \bibfield  {author} {\bibinfo {author} {\bibfnamefont {J.}~\bibnamefont
  {Dong}}, \bibinfo {author} {\bibfnamefont {Y.}~\bibnamefont {Hu}}, \bibinfo
  {author} {\bibfnamefont {J.}~\bibnamefont {Huang}}, \bibinfo {author}
  {\bibfnamefont {M.}~\bibnamefont {Ye}}, \bibinfo {author} {\bibfnamefont
  {Q.}~\bibnamefont {Qu}}, \bibinfo {author} {\bibfnamefont {T.}~\bibnamefont
  {Li}}, \ and\ \bibinfo {author} {\bibfnamefont {L.}~\bibnamefont {Liu}},\
  }\href@noop {} {\bibfield  {journal} {\bibinfo  {journal} {Applied optics}\
  }\textbf {\bibinfo {volume} {54}},\ \bibinfo {pages} {1152} (\bibinfo {year}
  {2015})}\BibitemShut {NoStop}%
\bibitem [{\citenamefont {Clerk}\ \emph {et~al.}(2010)\citenamefont {Clerk},
  \citenamefont {Devoret}, \citenamefont {Girvin}, \citenamefont {Marquardt},\
  and\ \citenamefont {Schoelkopf}}]{clerk2010introduction}%
  \BibitemOpen
  \bibfield  {author} {\bibinfo {author} {\bibfnamefont {A.~A.}\ \bibnamefont
  {Clerk}}, \bibinfo {author} {\bibfnamefont {M.~H.}\ \bibnamefont {Devoret}},
  \bibinfo {author} {\bibfnamefont {S.~M.}\ \bibnamefont {Girvin}}, \bibinfo
  {author} {\bibfnamefont {F.}~\bibnamefont {Marquardt}}, \ and\ \bibinfo
  {author} {\bibfnamefont {R.~J.}\ \bibnamefont {Schoelkopf}},\ }\href@noop {}
  {\bibfield  {journal} {\bibinfo  {journal} {Reviews of Modern Physics}\
  }\textbf {\bibinfo {volume} {82}},\ \bibinfo {pages} {1155} (\bibinfo {year}
  {2010})}\BibitemShut {NoStop}%
\bibitem [{\citenamefont {Kittel}(1988)}]{kittel1988temperature}%
  \BibitemOpen
  \bibfield  {author} {\bibinfo {author} {\bibfnamefont {C.}~\bibnamefont
  {Kittel}},\ }\href@noop {} {\bibfield  {journal} {\bibinfo  {journal} {Phys.
  Today}\ }\textbf {\bibinfo {volume} {41}},\ \bibinfo {pages} {93} (\bibinfo
  {year} {1988})}\BibitemShut {NoStop}%
\bibitem [{\citenamefont {Glenn}(1989)}]{glenn1989noise}%
  \BibitemOpen
  \bibfield  {author} {\bibinfo {author} {\bibfnamefont {W.~H.}\ \bibnamefont
  {Glenn}},\ }\href@noop {} {\bibfield  {journal} {\bibinfo  {journal} {IEEE
  journal of quantum electronics}\ }\textbf {\bibinfo {volume} {25}},\ \bibinfo
  {pages} {1218} (\bibinfo {year} {1989})}\BibitemShut {NoStop}%
\bibitem [{\citenamefont {Michaud-Belleau}\ \emph {et~al.}(2022)\citenamefont
  {Michaud-Belleau}, \citenamefont {Fokoua}, \citenamefont {Horak},
  \citenamefont {Wheeler}, \citenamefont {Rikimi}, \citenamefont {Bradley},
  \citenamefont {Richardson}, \citenamefont {Poletti}, \citenamefont {Genest},\
  and\ \citenamefont {Slav{\'\i}k}}]{michaud2022fundamental}%
  \BibitemOpen
  \bibfield  {author} {\bibinfo {author} {\bibfnamefont {V.}~\bibnamefont
  {Michaud-Belleau}}, \bibinfo {author} {\bibfnamefont {E.~R.~N.}\ \bibnamefont
  {Fokoua}}, \bibinfo {author} {\bibfnamefont {P.}~\bibnamefont {Horak}},
  \bibinfo {author} {\bibfnamefont {N.~V.}\ \bibnamefont {Wheeler}}, \bibinfo
  {author} {\bibfnamefont {S.}~\bibnamefont {Rikimi}}, \bibinfo {author}
  {\bibfnamefont {T.~D.}\ \bibnamefont {Bradley}}, \bibinfo {author}
  {\bibfnamefont {D.~J.}\ \bibnamefont {Richardson}}, \bibinfo {author}
  {\bibfnamefont {F.}~\bibnamefont {Poletti}}, \bibinfo {author} {\bibfnamefont
  {J.}~\bibnamefont {Genest}}, \ and\ \bibinfo {author} {\bibfnamefont
  {R.}~\bibnamefont {Slav{\'\i}k}},\ }\href@noop {} {\bibfield  {journal}
  {\bibinfo  {journal} {Physical Review A}\ }\textbf {\bibinfo {volume}
  {106}},\ \bibinfo {pages} {023501} (\bibinfo {year} {2022})}\BibitemShut
  {NoStop}%
\bibitem [{\citenamefont {Wanser}(1992)}]{wanser1992fundamental}%
  \BibitemOpen
  \bibfield  {author} {\bibinfo {author} {\bibfnamefont {K.}~\bibnamefont
  {Wanser}},\ }\href@noop {} {\bibfield  {journal} {\bibinfo  {journal}
  {Electronics letters}\ }\textbf {\bibinfo {volume} {28}},\ \bibinfo {pages}
  {53} (\bibinfo {year} {1992})}\BibitemShut {NoStop}%
\bibitem [{\citenamefont {Jasion}\ \emph {et~al.}(2020)\citenamefont {Jasion},
  \citenamefont {Bradley}, \citenamefont {Harrington}, \citenamefont {Sakr},
  \citenamefont {Chen}, \citenamefont {Fokoua}, \citenamefont {Davidson},
  \citenamefont {Taranta}, \citenamefont {Hayes}, \citenamefont {Richardson}
  \emph {et~al.}}]{jasion2020hollow}%
  \BibitemOpen
  \bibfield  {author} {\bibinfo {author} {\bibfnamefont {G.~T.}\ \bibnamefont
  {Jasion}}, \bibinfo {author} {\bibfnamefont {T.~D.}\ \bibnamefont {Bradley}},
  \bibinfo {author} {\bibfnamefont {K.}~\bibnamefont {Harrington}}, \bibinfo
  {author} {\bibfnamefont {H.}~\bibnamefont {Sakr}}, \bibinfo {author}
  {\bibfnamefont {Y.}~\bibnamefont {Chen}}, \bibinfo {author} {\bibfnamefont
  {E.~N.}\ \bibnamefont {Fokoua}}, \bibinfo {author} {\bibfnamefont {I.~A.}\
  \bibnamefont {Davidson}}, \bibinfo {author} {\bibfnamefont {A.}~\bibnamefont
  {Taranta}}, \bibinfo {author} {\bibfnamefont {J.~R.}\ \bibnamefont {Hayes}},
  \bibinfo {author} {\bibfnamefont {D.~J.}\ \bibnamefont {Richardson}},  \emph
  {et~al.},\ }in\ \href@noop {} {\emph {\bibinfo {booktitle} {Optical Fiber
  Communication Conference}}}\ (\bibinfo {organization} {Optical Society of
  America},\ \bibinfo {year} {2020})\ pp.\ \bibinfo {pages}
  {Th4B--4}\BibitemShut {NoStop}%
\bibitem [{\citenamefont {Butter}\ and\ \citenamefont
  {Hocker}(1978)}]{butter1978fiber}%
  \BibitemOpen
  \bibfield  {author} {\bibinfo {author} {\bibfnamefont {C.~D.}\ \bibnamefont
  {Butter}}\ and\ \bibinfo {author} {\bibfnamefont {G.}~\bibnamefont
  {Hocker}},\ }\href@noop {} {\bibfield  {journal} {\bibinfo  {journal}
  {Applied optics}\ }\textbf {\bibinfo {volume} {17}},\ \bibinfo {pages} {2867}
  (\bibinfo {year} {1978})}\BibitemShut {NoStop}%
\bibitem [{\citenamefont {Beadle}\ and\ \citenamefont
  {Jarzynski}(2001)}]{beadle2001measurement}%
  \BibitemOpen
  \bibfield  {author} {\bibinfo {author} {\bibfnamefont {B.~M.}\ \bibnamefont
  {Beadle}}\ and\ \bibinfo {author} {\bibfnamefont {J.}~\bibnamefont
  {Jarzynski}},\ }\href@noop {} {\bibfield  {journal} {\bibinfo  {journal}
  {Optical engineering}\ }\textbf {\bibinfo {volume} {40}},\ \bibinfo {pages}
  {2115} (\bibinfo {year} {2001})}\BibitemShut {NoStop}%
\bibitem [{\citenamefont {Schroeter}\ \emph {et~al.}(2007)\citenamefont
  {Schroeter}, \citenamefont {Nawrodt}, \citenamefont {Schnabel}, \citenamefont
  {Reid}, \citenamefont {Martin}, \citenamefont {Rowan}, \citenamefont
  {Schwarz}, \citenamefont {Koettig}, \citenamefont {Neubert}, \citenamefont
  {Th{\"u}rk} \emph {et~al.}}]{schroeter2007mechanical}%
  \BibitemOpen
  \bibfield  {author} {\bibinfo {author} {\bibfnamefont {A.}~\bibnamefont
  {Schroeter}}, \bibinfo {author} {\bibfnamefont {R.}~\bibnamefont {Nawrodt}},
  \bibinfo {author} {\bibfnamefont {R.}~\bibnamefont {Schnabel}}, \bibinfo
  {author} {\bibfnamefont {S.}~\bibnamefont {Reid}}, \bibinfo {author}
  {\bibfnamefont {I.}~\bibnamefont {Martin}}, \bibinfo {author} {\bibfnamefont
  {S.}~\bibnamefont {Rowan}}, \bibinfo {author} {\bibfnamefont
  {C.}~\bibnamefont {Schwarz}}, \bibinfo {author} {\bibfnamefont
  {T.}~\bibnamefont {Koettig}}, \bibinfo {author} {\bibfnamefont
  {R.}~\bibnamefont {Neubert}}, \bibinfo {author} {\bibfnamefont
  {M.}~\bibnamefont {Th{\"u}rk}},  \emph {et~al.},\ }\href@noop {} {\bibfield
  {journal} {\bibinfo  {journal} {arXiv preprint arXiv:0709.4359}\ } (\bibinfo
  {year} {2007})}\BibitemShut {NoStop}%
\bibitem [{\citenamefont {Gretarsson}\ and\ \citenamefont
  {Harry}(1999)}]{gretarsson1999dissipation}%
  \BibitemOpen
  \bibfield  {author} {\bibinfo {author} {\bibfnamefont {A.~M.}\ \bibnamefont
  {Gretarsson}}\ and\ \bibinfo {author} {\bibfnamefont {G.~M.}\ \bibnamefont
  {Harry}},\ }\href@noop {} {\bibfield  {journal} {\bibinfo  {journal} {Review
  of scientific instruments}\ }\textbf {\bibinfo {volume} {70}},\ \bibinfo
  {pages} {4081} (\bibinfo {year} {1999})}\BibitemShut {NoStop}%
\bibitem [{\citenamefont {Penn}\ \emph {et~al.}(2001)\citenamefont {Penn},
  \citenamefont {Harry}, \citenamefont {Gretarsson}, \citenamefont
  {Kittelberger}, \citenamefont {Saulson}, \citenamefont {Schiller},
  \citenamefont {Smith},\ and\ \citenamefont {Swords}}]{penn2001high}%
  \BibitemOpen
  \bibfield  {author} {\bibinfo {author} {\bibfnamefont {S.~D.}\ \bibnamefont
  {Penn}}, \bibinfo {author} {\bibfnamefont {G.~M.}\ \bibnamefont {Harry}},
  \bibinfo {author} {\bibfnamefont {A.~M.}\ \bibnamefont {Gretarsson}},
  \bibinfo {author} {\bibfnamefont {S.~E.}\ \bibnamefont {Kittelberger}},
  \bibinfo {author} {\bibfnamefont {P.~R.}\ \bibnamefont {Saulson}}, \bibinfo
  {author} {\bibfnamefont {J.~J.}\ \bibnamefont {Schiller}}, \bibinfo {author}
  {\bibfnamefont {J.~R.}\ \bibnamefont {Smith}}, \ and\ \bibinfo {author}
  {\bibfnamefont {S.~O.}\ \bibnamefont {Swords}},\ }\href@noop {} {\bibfield
  {journal} {\bibinfo  {journal} {Review of Scientific Instruments}\ }\textbf
  {\bibinfo {volume} {72}},\ \bibinfo {pages} {3670} (\bibinfo {year}
  {2001})}\BibitemShut {NoStop}%
\bibitem [{\citenamefont {Heptonstall}\ \emph {et~al.}(2010)\citenamefont
  {Heptonstall}, \citenamefont {Barton}, \citenamefont {Cantley}, \citenamefont
  {Cumming}, \citenamefont {Cagnoli}, \citenamefont {Hough}, \citenamefont
  {Jones}, \citenamefont {Kumar}, \citenamefont {Martin}, \citenamefont {Rowan}
  \emph {et~al.}}]{heptonstall2010investigation}%
  \BibitemOpen
  \bibfield  {author} {\bibinfo {author} {\bibfnamefont {A.}~\bibnamefont
  {Heptonstall}}, \bibinfo {author} {\bibfnamefont {M.}~\bibnamefont {Barton}},
  \bibinfo {author} {\bibfnamefont {C.}~\bibnamefont {Cantley}}, \bibinfo
  {author} {\bibfnamefont {A.}~\bibnamefont {Cumming}}, \bibinfo {author}
  {\bibfnamefont {G.}~\bibnamefont {Cagnoli}}, \bibinfo {author} {\bibfnamefont
  {J.}~\bibnamefont {Hough}}, \bibinfo {author} {\bibfnamefont
  {R.}~\bibnamefont {Jones}}, \bibinfo {author} {\bibfnamefont
  {R.}~\bibnamefont {Kumar}}, \bibinfo {author} {\bibfnamefont
  {I.}~\bibnamefont {Martin}}, \bibinfo {author} {\bibfnamefont
  {S.}~\bibnamefont {Rowan}},  \emph {et~al.},\ }\href@noop {} {\bibfield
  {journal} {\bibinfo  {journal} {Classical and Quantum Gravity}\ }\textbf
  {\bibinfo {volume} {27}},\ \bibinfo {pages} {035013} (\bibinfo {year}
  {2010})}\BibitemShut {NoStop}%
\bibitem [{\citenamefont {McSkimin}(1953)}]{mcskimin1953measurement}%
  \BibitemOpen
  \bibfield  {author} {\bibinfo {author} {\bibfnamefont {H.}~\bibnamefont
  {McSkimin}},\ }\href@noop {} {\bibfield  {journal} {\bibinfo  {journal}
  {Journal of applied physics}\ }\textbf {\bibinfo {volume} {24}},\ \bibinfo
  {pages} {988} (\bibinfo {year} {1953})}\BibitemShut {NoStop}%
\bibitem [{\citenamefont {Reid}\ and\ \citenamefont
  {Ozcan}(1998)}]{reid1998temperature}%
  \BibitemOpen
  \bibfield  {author} {\bibinfo {author} {\bibfnamefont {M.~B.}\ \bibnamefont
  {Reid}}\ and\ \bibinfo {author} {\bibfnamefont {M.}~\bibnamefont {Ozcan}},\
  }\href@noop {} {\bibfield  {journal} {\bibinfo  {journal} {Optical
  Engineering}\ }\textbf {\bibinfo {volume} {37}},\ \bibinfo {pages} {237}
  (\bibinfo {year} {1998})}\BibitemShut {NoStop}%
\bibitem [{\citenamefont {Behunin}\ \emph {et~al.}(2017)\citenamefont
  {Behunin}, \citenamefont {Kharel}, \citenamefont {Renninger},\ and\
  \citenamefont {Rakich}}]{behunin2017engineering}%
  \BibitemOpen
  \bibfield  {author} {\bibinfo {author} {\bibfnamefont {R.}~\bibnamefont
  {Behunin}}, \bibinfo {author} {\bibfnamefont {P.}~\bibnamefont {Kharel}},
  \bibinfo {author} {\bibfnamefont {W.}~\bibnamefont {Renninger}}, \ and\
  \bibinfo {author} {\bibfnamefont {P.}~\bibnamefont {Rakich}},\ }\href@noop {}
  {\bibfield  {journal} {\bibinfo  {journal} {Nature Materials}\ }\textbf
  {\bibinfo {volume} {16}},\ \bibinfo {pages} {315} (\bibinfo {year}
  {2017})}\BibitemShut {NoStop}%
\bibitem [{\citenamefont {Zhu}\ \emph {et~al.}(2019)\citenamefont {Zhu},
  \citenamefont {Fokoua}, \citenamefont {Taranta}, \citenamefont {Chen},
  \citenamefont {Bradley}, \citenamefont {Petrovich}, \citenamefont {Poletti},
  \citenamefont {Zhao}, \citenamefont {Richardson},\ and\ \citenamefont
  {Slav{\'\i}k}}]{zhu2019thermal}%
  \BibitemOpen
  \bibfield  {author} {\bibinfo {author} {\bibfnamefont {W.}~\bibnamefont
  {Zhu}}, \bibinfo {author} {\bibfnamefont {E.~R.~N.}\ \bibnamefont {Fokoua}},
  \bibinfo {author} {\bibfnamefont {A.~A.}\ \bibnamefont {Taranta}}, \bibinfo
  {author} {\bibfnamefont {Y.}~\bibnamefont {Chen}}, \bibinfo {author}
  {\bibfnamefont {T.}~\bibnamefont {Bradley}}, \bibinfo {author} {\bibfnamefont
  {M.~N.}\ \bibnamefont {Petrovich}}, \bibinfo {author} {\bibfnamefont
  {F.}~\bibnamefont {Poletti}}, \bibinfo {author} {\bibfnamefont
  {M.}~\bibnamefont {Zhao}}, \bibinfo {author} {\bibfnamefont {D.~J.}\
  \bibnamefont {Richardson}}, \ and\ \bibinfo {author} {\bibfnamefont
  {R.}~\bibnamefont {Slav{\'\i}k}},\ }\href@noop {} {\bibfield  {journal}
  {\bibinfo  {journal} {Journal of Lightwave Technology}\ }\textbf {\bibinfo
  {volume} {38}},\ \bibinfo {pages} {2477} (\bibinfo {year}
  {2019})}\BibitemShut {NoStop}%
\bibitem [{\citenamefont {Huang}\ \emph
  {et~al.}(2019{\natexlab{b}})\citenamefont {Huang}, \citenamefont {Wang},
  \citenamefont {Duan}, \citenamefont {Huang}, \citenamefont {Ye},
  \citenamefont {Li}, \citenamefont {Liu},\ and\ \citenamefont
  {Li}}]{huang2019vibration}%
  \BibitemOpen
  \bibfield  {author} {\bibinfo {author} {\bibfnamefont {J.}~\bibnamefont
  {Huang}}, \bibinfo {author} {\bibfnamefont {L.}~\bibnamefont {Wang}},
  \bibinfo {author} {\bibfnamefont {Y.}~\bibnamefont {Duan}}, \bibinfo {author}
  {\bibfnamefont {Y.}~\bibnamefont {Huang}}, \bibinfo {author} {\bibfnamefont
  {M.}~\bibnamefont {Ye}}, \bibinfo {author} {\bibfnamefont {L.}~\bibnamefont
  {Li}}, \bibinfo {author} {\bibfnamefont {L.}~\bibnamefont {Liu}}, \ and\
  \bibinfo {author} {\bibfnamefont {T.}~\bibnamefont {Li}},\ }\href@noop {}
  {\bibfield  {journal} {\bibinfo  {journal} {Chinese Optics Letters}\ }\textbf
  {\bibinfo {volume} {17}},\ \bibinfo {pages} {081403} (\bibinfo {year}
  {2019}{\natexlab{b}})}\BibitemShut {NoStop}%
\bibitem [{\citenamefont {Thorpe}\ \emph {et~al.}(2010)\citenamefont {Thorpe},
  \citenamefont {Leibrandt}, \citenamefont {Fortier},\ and\ \citenamefont
  {Rosenband}}]{thorpe2010measurement}%
  \BibitemOpen
  \bibfield  {author} {\bibinfo {author} {\bibfnamefont {M.~J.}\ \bibnamefont
  {Thorpe}}, \bibinfo {author} {\bibfnamefont {D.~R.}\ \bibnamefont
  {Leibrandt}}, \bibinfo {author} {\bibfnamefont {T.~M.}\ \bibnamefont
  {Fortier}}, \ and\ \bibinfo {author} {\bibfnamefont {T.}~\bibnamefont
  {Rosenband}},\ }\href@noop {} {\bibfield  {journal} {\bibinfo  {journal}
  {Optics Express}\ }\textbf {\bibinfo {volume} {18}},\ \bibinfo {pages}
  {18744} (\bibinfo {year} {2010})}\BibitemShut {NoStop}%
\bibitem [{\citenamefont {Leibrandt}\ \emph {et~al.}(2013)\citenamefont
  {Leibrandt}, \citenamefont {Bergquist},\ and\ \citenamefont
  {Rosenband}}]{leibrandt2013cavity}%
  \BibitemOpen
  \bibfield  {author} {\bibinfo {author} {\bibfnamefont {D.~R.}\ \bibnamefont
  {Leibrandt}}, \bibinfo {author} {\bibfnamefont {J.~C.}\ \bibnamefont
  {Bergquist}}, \ and\ \bibinfo {author} {\bibfnamefont {T.}~\bibnamefont
  {Rosenband}},\ }\href@noop {} {\bibfield  {journal} {\bibinfo  {journal}
  {Physical Review A}\ }\textbf {\bibinfo {volume} {87}},\ \bibinfo {pages}
  {023829} (\bibinfo {year} {2013})}\BibitemShut {NoStop}%
\bibitem [{\citenamefont {Proakis}\ and\ \citenamefont
  {Manolakis}(2006)}]{DSPBook}%
  \BibitemOpen
  \bibfield  {author} {\bibinfo {author} {\bibfnamefont {J.~G.}\ \bibnamefont
  {Proakis}}\ and\ \bibinfo {author} {\bibfnamefont {D.~K.}\ \bibnamefont
  {Manolakis}},\ }\href@noop {} {\emph {\bibinfo {title} {Digital Signal
  Processing (4th Edition)}}},\ \bibinfo {edition} {4th}\ ed.\ (\bibinfo
  {publisher} {Prentice Hall},\ \bibinfo {year} {2006})\BibitemShut {NoStop}%
\bibitem [{\citenamefont {Beuermann}\ \emph {et~al.}(1999)\citenamefont
  {Beuermann}, \citenamefont {Hessman}, \citenamefont {Reinsch}, \citenamefont
  {Nicklas}, \citenamefont {Vreeswijk}, \citenamefont {Galama}, \citenamefont
  {Rol}, \citenamefont {van Paradijs}, \citenamefont {Kouveliotou},
  \citenamefont {Frontera}, \citenamefont {Masetti}, \citenamefont {Palazzi},\
  and\ \citenamefont {Pian}}]{brokenPowerLaw}%
  \BibitemOpen
  \bibfield  {author} {\bibinfo {author} {\bibfnamefont {K.}~\bibnamefont
  {Beuermann}}, \bibinfo {author} {\bibfnamefont {F.~V.}\ \bibnamefont
  {Hessman}}, \bibinfo {author} {\bibfnamefont {K.}~\bibnamefont {Reinsch}},
  \bibinfo {author} {\bibfnamefont {H.}~\bibnamefont {Nicklas}}, \bibinfo
  {author} {\bibfnamefont {P.~M.}\ \bibnamefont {Vreeswijk}}, \bibinfo {author}
  {\bibfnamefont {T.~J.}\ \bibnamefont {Galama}}, \bibinfo {author}
  {\bibfnamefont {E.}~\bibnamefont {Rol}}, \bibinfo {author} {\bibfnamefont
  {J.}~\bibnamefont {van Paradijs}}, \bibinfo {author} {\bibfnamefont
  {C.}~\bibnamefont {Kouveliotou}}, \bibinfo {author} {\bibfnamefont
  {F.}~\bibnamefont {Frontera}}, \bibinfo {author} {\bibfnamefont
  {N.}~\bibnamefont {Masetti}}, \bibinfo {author} {\bibfnamefont
  {E.}~\bibnamefont {Palazzi}}, \ and\ \bibinfo {author} {\bibfnamefont
  {E.}~\bibnamefont {Pian}},\ }\href@noop {} {\enquote {\bibinfo {title} {Vlt
  observations of grb 990510 and its environment},}\ } (\bibinfo {year}
  {1999}),\ \Eprint {http://arxiv.org/abs/astro-ph/9909043}
  {arXiv:astro-ph/9909043 [astro-ph]} \BibitemShut {NoStop}%
\end{thebibliography}%

\end{document}